\documentclass[runninghead]{llncs}

\usepackage{amsmath,amsfonts,amssymb}

\usepackage{fancyhdr} 
\usepackage{lipsum}
\usepackage{tikz}
\usetikzlibrary{arrows.meta}
\usetikzlibrary{automata,positioning}
\usepackage{latexsym}
\usepackage{float}

\usepackage{fbox} 
\usepackage{fancybox}
\usepackage{framed}

\usepackage{lscape}
\usepackage{rotating}
\usepackage[graphicx]{realboxes}
\usepackage{adjustbox}
\usepackage{dashbox}
\usepackage{etoolbox}
\usepackage[most]{tcolorbox}

\usepackage{mathtools}

\usepackage{booktabs} 

\usepackage{listings}
\usepackage{blindtext}
\usepackage[ruled,linesnumbered]{algorithm2e}
\usepackage{etoolbox,xstring,mfirstuc,textcase}
\usepackage{booktabs} 
\usepackage{listings}

\usepackage{xcolor}

\usepackage{indentfirst} 

\usepackage{dashbox}
\usepackage{amsfonts}
\usepackage{booktabs}
\usepackage{siunitx}
\usepackage{multirow}
\usepackage{comment}
\usepackage{indentfirst} 
\usepackage{framed} 

\usepackage[font=small,labelfont=bf,tableposition=top]{caption}
\usepackage{booktabs}
\usepackage{threeparttable}
\usepackage{dashbox}
\usepackage{amsfonts}

\usepackage{cases}

\usepackage{graphicx}
\usepackage[lofdepth,lotdepth]{subfig}

\usepackage{lipsum}

\usepackage{tikz} 
\usepackage{comment}
\usepackage{filecontents}
\usepackage{url}
\usepackage{hyperref}
\usepackage{multirow}

\usepackage{algorithmic}
\usepackage{graphicx}
\usepackage{textcomp}
\usepackage{xcolor}
\usepackage{amsthm}
\usepackage{booktabs}
\usepackage[utf8]{inputenc}

\usepackage{dashbox}
\usepackage{todonotes}
\usepackage{enumitem}   
\usetikzlibrary{shapes.callouts}
\usepackage{listings}
\usepackage{trace}
\usepackage{mdframed}

\theoremstyle{plain}
 
\newtheorem{thm}{Theorem}   
  
\newtheorem{defi}{Definition}
\theoremstyle{definition}

\newtheorem{prf}{Proof}

\usepackage[
	n,
	operators,
	advantage,
	sets,
	adversary,
	landau,
	probability,
	notions,	
	logic,
	ff,
	mm,
	primitives,
	events,
	complexity,
	asymptotics,
	keys]{cryptocode}
	
\usepackage{dashbox}
\usepackage{tikz}

\usepackage{ulem}

\usepackage{tikz}

\usetikzlibrary{arrows, calc, decorations.markings, positioning}

\makeatletter
\newenvironment{timeline}[6]{%

    \newcommand{\startyear}{#1}
    \newcommand{\tlendyear}{#2}

    \newcommand{\yearcolumnwidth}{#3}
    \newcommand{\rulecolumnwidth}{#4}
    \newcommand{\entrycolumnwidth}{#5}
    \newcommand{\timelineheight}{#6}

    \newcommand{\templength}{}

    \newcommand{\entrycounter}{0}

    \long\def\ifnodedefined##1##2##3{%
        \@ifundefined{pgf@sh@ns@##1}{##3}{##2}%
    }

    \newcommand{\ifnodeundefined}[2]{%
        \ifnodedefined{##1}{}{##2}
    }

    \newcommand{\drawtimeline}{%
        \draw[timelinerule] (\yearcolumnwidth+5pt, 0pt) -- (\yearcolumnwidth+5pt, -\timelineheight);
        \draw (\yearcolumnwidth+0pt, -10pt) -- (\yearcolumnwidth+10pt, -10pt);
        \draw (\yearcolumnwidth+0pt, -\timelineheight+15pt) -- (\yearcolumnwidth+10pt, -\timelineheight+15pt);

        \pgfmathsetlengthmacro{\templength}{neg(add(multiply(subtract(\startyear, \startyear), divide(subtract(\timelineheight, 25), subtract(\tlendyear, \startyear))), 10))}
        \node[year] (year-\startyear) at (\yearcolumnwidth, \templength) {\startyear};

        \pgfmathsetlengthmacro{\templength}{neg(add(multiply(subtract(\tlendyear, \startyear), divide(subtract(\timelineheight, 25), subtract(\tlendyear, \startyear))), 10))}
        \node[year] (year-\tlendyear) at (\yearcolumnwidth, \templength) {\tlendyear};
    }

    \newcommand{\entry}[2]{%

        \pgfmathtruncatemacro{\lastentrycount}{\entrycounter}
        \pgfmathtruncatemacro{\entrycounter}{\entrycounter + 1}

        \ifdim \lastentrycount pt > 0 pt%
            \node[entry] (entry-\entrycounter) [below of=entry-\lastentrycount] {##2};
        \else%
            \pgfmathsetlengthmacro{\templength}{neg(add(multiply(subtract(\startyear, \startyear), divide(subtract(\timelineheight, 25), subtract(\tlendyear, \startyear))), 10))}
            \node[entry] (entry-\entrycounter) at (\yearcolumnwidth+\rulecolumnwidth+10pt, \templength) {##2};
        \fi

        \ifnodeundefined{year-##1}{%
            \pgfmathsetlengthmacro{\templength}{neg(add(multiply(subtract(##1, \startyear), divide(subtract(\timelineheight, 25), subtract(\tlendyear, \startyear))), 10))}
            \draw (\yearcolumnwidth+2.5pt, \templength) -- (\yearcolumnwidth+7.5pt, \templength);
            \node[year] (year-##1) at (\yearcolumnwidth, \templength) {##1};
        }

        \draw ($(year-##1.east)+(2.5pt, 0pt)$) -- ($(year-##1.east)+(7.5pt, 0pt)$) -- ($(entry-\entrycounter.west)-(5pt,0)$) -- (entry-\entrycounter.west);
    }

    \newcommand{\plainentry}[2]{

        \pgfmathtruncatemacro{\lastentrycount}{\entrycounter}
        \pgfmathtruncatemacro{\entrycounter}{\entrycounter + 1}

        \ifdim \lastentrycount pt > 0 pt%
            \node[entry] (entry-\entrycounter) [below of=entry-\lastentrycount] {##2};
        \else%
            \pgfmathsetlengthmacro{\templength}{neg(add(multiply(subtract(\startyear, \startyear), divide(subtract(\timelineheight, 25), subtract(\tlendyear, \startyear))), 10))}
            \node[entry] (entry-\entrycounter) at (\yearcolumnwidth+\rulecolumnwidth+10pt, \templength) {##2};
        \fi

        \ifnodeundefined{invisible-year-##1}{%
            \pgfmathsetlengthmacro{\templength}{neg(add(multiply(subtract(##1, \startyear), divide(subtract(\timelineheight, 25), subtract(\tlendyear, \startyear))), 10))}
            \draw (\yearcolumnwidth+2.5pt, \templength) -- (\yearcolumnwidth+7.5pt, \templength);
            \node[year] (invisible-year-##1) at (\yearcolumnwidth, \templength) {};
        }

        \draw ($(invisible-year-##1.east)+(2.5pt, 0pt)$) -- ($(invisible-year-##1.east)+(7.5pt, 0pt)$) -- ($(entry-\entrycounter.west)-(5pt,0)$) -- (entry-\entrycounter.west);
    }

    \begin{tikzpicture}
        \tikzstyle{entry} = [%
            align=left,%
            text width=\entrycolumnwidth,%
            node distance=10mm,%
            anchor=west]
        \tikzstyle{year} = [anchor=east]
        \tikzstyle{timelinerule} = [%
            draw,%
            decoration={markings, mark=at position 1 with {\arrow[scale=1.5]{latex'}}},%
            postaction={decorate},%
            shorten >=0.4pt]

        \drawtimeline
}
{
    \end{tikzpicture}
    \let\startyear\@undefined
    \let\tlendyear\@undefined
    \let\yearcolumnwidth\@undefined
    \let\rulecolumnwidth\@undefined
    \let\entrycolumnwidth\@undefined
    \let\timelineheight\@undefined
    \let\entrycounter\@undefined
    \let\ifnodedefined\@undefined
    \let\ifnodeundefined\@undefined
    \let\drawtimeline\@undefined
    \let\entry\@undefined
}
\makeatother

\usetikzlibrary{shapes,arrows}

\pgfdeclarelayer{background}
\pgfdeclarelayer{foreground}
\pgfsetlayers{background,main,foreground}

\tikzstyle{sensor}=[draw, fill=blue!20, text width=5em, 
    text centered, minimum height=2.5em]
\tikzstyle{ann} = [above, text width=3.5em]
\tikzstyle{naveqs} = [sensor, text width=6em, fill=red!20, 
    minimum height=12em, rounded corners]
\def\blockdist{2.0}
\def\edgedist{1.8}

\begin{document}

\title{How Do Smart Contracts Benefit \\Security Protocols?}


\author{Rujia Li$^{\star}$\inst{1,2}, Qin Wang$\thanks{These authors contributed equally to the work.\\ Email: \url{rxl635@bham.ac.uk} and \url{qinwangtech@gmail.com}.}$\inst{3}, Qi Wang\inst{1}, David Galindo\inst{2}}
\authorrunning{Qin Wang}
\authorrunning{Rujia Li}
\titlerunning{Smart Contract Benefits}

\institute{
\textit{Southern University of Science and Technology}, China
\\
\and
\textit{University of Birmingham}, UK
\\
\and
\textit{CSIRO Data61}, Australia
\\
}


\maketitle

\begin{abstract}
Smart contracts have recently been adopted by many security protocols. However, existing studies lack satisfactory theoretical support on how contracts benefit security protocols. This paper aims to give a systematic analysis of smart contract (SC)-based security protocols to fulfill the gap of unclear arguments and statements. We firstly investigate \textit{state of the art studies} and establish a formalized model of smart contract protocols with well-defined syntax and assumptions. Then, we apply our formal framework to two concrete instructions to explore corresponding advantages and desirable properties. Through our analysis, we abstract three generic properties (\textit{non-repudiation, non-equivocation, and non-frameability}) and accordingly identify two patterns. (1) a smart contract can be as an autonomous subscriber to assist the trusted third party (TTP); (2) a smart contract can replace traditional TTP. To the best of our knowledge, this is the first study to provide in-depth discussions of SC-based security protocols from a strictly theoretical perspective.

%
\end{abstract}



\section{Introduction}

The smart contract (SC) was initially introduced by Szabo~\cite{szabo1996smart} who suggests that the clauses of a contract should be self-executed by the ways like being translated into code and embedded into software to minimize contracting cost between transacting parties as well as avoiding accidental exceptions or malicious actions. Such an idea has been developed with the advent of blockchain systems. Ethereum~\cite{wood2014ethereum} implemented the first practical smart contract with persistent data storage on distributed ledgers, enabling the support and management of complete lifecycles of legal contracts. The creation of documents and the subsequent use of those templates by counterparties becomes feasible in large-scale cooperation. Technically, Ethereum-based smart contract utilizes Turing-complete scripting languages to achieve complicated functionalities~\cite{jansen2019smart} and execute thorough state transition/replication over consensus algorithms for final consistency. Based on that, the contracts running in a distributed network can still be checked publicly and traced by all participants. 


The smart contracts have been employed as cryptographic building blocks in many cryptographic protocols, such as PKI systems \cite{wang2018blockchain,dong2018conifer}, transparent log systems \cite{bonneau2016ethiks,melara2015coniks}, certificateless systems \cite{ali2019blockchain,eltayieb2019certificateless}, web of trust  \cite{roh2018study,wilson2015pretty,dunphy2018first}, etc. The smart contracts make the operations of a service provider transparent: all loaded data to the contract and corresponding operations are publicly checkable and traceable. For example, CONIKS~\cite{bonneau2016ethiks} proposed an auditable key transparency system under the help of the Ethereum-based smart contract. Users can audit the identity-key binding relationship by checking the unambiguous state in the contract. Similarly, Wang et al. \cite{wang2018blockchain} utilized a smart contract to balance the absolute authority of certificate authorities (CAs), implementing the certificate transparency and revocation transparency. The smart contract helps to construct public logs, where a CA-signed certificate has to be published by certificate transactions in the global blockchain. The scheme, thus, significantly reinforces the security guarantees of a certificate.

Although plenty of solutions have adopted the smart contract technique, a comprehensive analysis on \textit{how to combine smart contracts with security protocols} and \textit{what types of advantages can be obtained} is absent. Establishing a generic security analysis framework is a non-trivial task and needs a lot of efforts. Some studies, such as \cite{locher2018can}, merely provide preliminary discussions without solid evidence or strict models. This paper aims to dive into such problems by abstracting a formal and generic framework of the smart contract, exploring the obtained benefits when applying it to security protocols. We provide security definitions as of the corresponding syntax. Then, we apply our formalized framework to two distinguished instances, including the certificate-based encryption (CBE) scheme~\cite{wang2018blockchain,bonneau2016ethiks} and registration-based encryption (RBE) scheme~\cite{RBE18,RBE19}. These examples show a proper usage of smart contracts in terms of traditional security protocols. Besides, our formalization further lays foundations for extensible contract-based protocols such as DeFi~\cite{werner2021sok} and NFTs \cite{wang2021non}, which hold billions of market cap (measured by US dollars). In summary, this paper contributes in the following aspects.

\begin{itemize}

\item[-]  We review \textit{state-of-the-art} studies that leverage smart contracts in their security protocols. Based on comprehensive investigations, we capture generic features (covering design patterns and properties) that are insightful in a variety of scenarios. 

\item[-] We identify the benefits of smart contracts when combining them with traditional security protocols. We answer the questions of \textit{how smart contract combined with these protocols} and \textit{what benefits can be further obtained}. For the former, we identify two types of usages in terms of smart contracts in the context of security protocols: \textit{indirectly as a subscriber or public bulletin that assists the existing TTP} and \textit{directly as an agent that replaces TTP}. For the latter, we observe that smart contract brings security protocols with advanced properties of \textit{transparency}, \textit{decentralization} and \textit{accountability}.

\item[-] We formalize a smart contract with a rigorous blockchain model. The formalized framework is abstracted from Ethereum (both literature and implementations \cite{wood2014ethereum}) with extreme simplicity as well as capturing all the key features of the smart contract. It fits a variety of scenarios when applied to security protocols. The blockchain model is based on the assumptions of a robust public ledger that holds \textit{persistence} and \textit{liveness}~\cite{garay2015bitcoin}.

\item[-] A uniform framework of the smart contract-based protocol is provided, and its security properties are defined. These security properties include \textit{non-equivocation}, \textit{non-repudiation} and \textit{non-frameability}. We argue that the hybrid protocols combined with smart contracts enjoy (at least one of) these security properties.

\item[-] We apply the proposed framework to two types of cryptographic instances that cover the CBE scheme \cite{wang2018blockchain,bonneau2016ethiks} and the RBE scheme  \cite{garg2019registration,garg2018registration}. We present these protocols with detailed explanations of their combination steps and provide strict proofs of their properties. We demonstrate that our generalized formal framework is feasible from the theoretical view. We further provide in-depth discussions of the benefits and challenges existing in current solutions. 
\end{itemize}


By doing so, we observe that smart contracts enhance security protocols in two ways: assist existing TTP and replace the existing TTP. The first approach is to assist existing TTP with adding the functionalities of public access and automated records. Our hybrid CBE scheme employs its transparency and audibility properties to make each certificate revocation reliable. The second approach is to replace the existing TTP. Our transparent RBE scheme utilizes the smart contract as a Key Curator to manage the registration information of users. With the help of smart contracts, security protocols get improved by arming with (publicly) accessible states, transparent executions and accountable behaviours towards participants.

\smallskip
\noindent\textbf{Paper structure}. Section \ref{sec-relatedwk} gives the related studies. Section \ref{sec-model} presents the blockchain assumption and a formal treatment of smart contracts. Section \ref{sec-secureprotocol} provides the hybrid smart contract-based security protocol with corresponding properties. Section~\ref{sec-instance} demonstrates the feasibility of our framework by providing two typical examples, followed by their security proofs in Section~\ref{sec-proof}. Further discussions and experiment results are provided in Section \ref{sec-discussion}. Finally, Section \ref{sec-conclu} gives concluding remarks. 

\section{Related Work}
\label{sec-relatedwk}

This section provides concurrent studies from twofold: the formal treatment towards blockchain systems and a bird view of the smart contract (SC)-based security protocols.

\smallskip
\noindent\textbf{Formal Treatment.} 
Despite formally modelling blockchain systems is not a trivial task for users, many researchers have spared their great efforts in independent lines. Garay et al.~\cite{garay2015bitcoin} proposed the first abstraction of blockchain protocol. They have formally extracted two intrinsic properties in both static settings \cite{garay2015bitcoin} and dynamic settings \cite{garay2017bitcoin}. Fitzi et al. \cite{fitzi2018parallel} introduced a formal execution model of the parallel-chains paradigm. The model expresses transaction throughput as well as supporting formal security arguments of its basic properties (safety and liveness). Wang et al. \cite{wang2021formal} utilized the state machine replica (SMR) model to analyse a BFT-style consensus protocol. They formally discuss the insecurity of an improper modification of consensus mechanisms. Poulami \cite{das2019formal} provided a formal security model to analyse hot/cold wallets. They explored the security properties that wallets should hold. Andrew et al. \cite{lewis2021does} formally analyzed the factors (especially, \textit{certificates}) that determine the security of permissionless protocols. Furthermore, they clarified subtle differences in terms of security notions (live/safe/adaptive), committee settings (sized/unsized), user selection (PoW/PoS) and network assumptions (synchronous/partially synchronous). Besides, the formal treatment of system privacy or security \cite{camenisch2015formal} also provides many insights for this paper. 

\smallskip
\noindent\textbf{SC-based Security Protocols.} 
Security protocols confront issues caused by centralization, such as single-point failure in PKI solutions, key escrow problems in IBE schemes, etc. Blockchain-based smart contracts mitigate such issues in hybrid protocols (security protocols equipped with SC) by offering decentralization and accountability.  Hybrid protocols cover many subsets of traditional security protocols including PKI \cite{matsumoto2017ikp,qin2017cecoin,patsonakis2017towards}, transparency log \cite{wang2018blockchain,bonneau2016ethiks,wang2020blockchain}, web of trust \cite{roh2018study,wilson2015pretty,durand2017decentralized}, name service \cite{ali2016blockstack,xia2021ethereum}, identity management \cite{dunphy2018first,lim2018blockchain,liu2017identity}, certificateless encryption \cite{ali2019blockchain,rujia2020poster}, and registration-based encryption \cite{RBE18}. These studies either use the smart contract as a transparent bulletin board for public accessibility (e.g., PKI, web of trust), or leverage it as the data manager to register, merge, revoke or record the operations of stored keys/identifies (CBE, RBE). However, formal treatment and strict analysis of these solutions are absent. To fill the gap, this paper abstracts a generic model, as the theoretical support, to shows how smart contracts practically assist with such hybrid security protocols. 


\section{Blockchain and Smart Contract Formalization}
\label{sec-model}

In this section, we first provide a simple blockchain model, emphasizing its \textit{persistence} and \textit{liveness} assumption~\cite{garay2015bitcoin}. Then, based on that, we give a formal treatment of smart contract and contract-based protocols.

\subsection{Blockchain System}
Blockchain is a distributed and append-only ledger that drives from Bitcoin~\cite{nakamoto2008bitcoin}. Thus, for simplicity, we define a blockchain as a distributed database $\mathbb{B}$ with the following functionalities.

\begin{itemize} 

\item[-]  $\mathsf{(Tx, state_{i+1}) \gets \mathbb{B}.Write(Tx,state_{i})}.$ The blockchain players (also called miners or maintainers) write transactions to the database with updating the on-chain state. ($\mathsf{i \leq Total(player)}$, where $\mathsf{ Total(player)}$ represents the total number of the players.)

\item[-]  $\mathsf{state \gets \mathbb{B}.ReadState(Tx)}.$ Given a transaction $\mathsf{Tx}$, a user reads the confirmed state from any players. 

\item[-]  $\mathsf{Tx \gets \mathbb{B}.ReadTx(state)}.$ Given a confirmed state, a user finds the transaction that triggers the execution of state updating.

\end{itemize}

To capture security assumptions of a robust blockchain system, we also define the following expressions.

\begin{itemize}

\item[-] $\mathsf{List(player) \gets FindPlayer(Tx,k,\delta)}.$ It is used to find the blockchain players who have accepted the transaction (stored a transaction more than $k$ blocks deep in its local ledger) in a time-bound $\delta$.

\item[-] $\mathsf{Tx \gets FindTx(player_i,k)}$ It is used to find the transaction that a player has accepted more than $k$ blocks deep.

\end{itemize}

\begin{defi}[\textbf{Blockchain Assumption \cite{garay2015bitcoin}}] {A blockchain is a robust public transaction ledger if it satisfies the following properties.}
\end{defi}
    
\begin{itemize}
    \item[-] \textit{\textbf{Persistence.} Once one honest player stores a transaction more than $k$ blocks deep into its local ledger, other honest players (the number depending on a certain consensus algorithm) will reject such a transaction with a negligible probability, namely, $\mathsf{adv_{\adv,(k,\delta)}^{persistence} = }$}
    
\begin{center}
\begin{align*}
   \mathsf{Pr}\left[
    \begin{array}{ll}
     \mathsf{Tx \gets FindTx(player_i,k);}\\
     \mathsf{List(player) \gets FindPlayer(Tx,k,\delta);}\\
     \mathsf{Num(List(player))/Total(player)  \leq  \mathsf{\epsilon}  }
   \end{array}
   \right]  \leq \mathsf{negl(\lambda)},
\end{align*}
\end{center}

\textit{where $\mathsf{player_i}$ represents any honest player, $\mathsf{Num(List(player))}$ means the number of the blockchain players who have accepted $\mathsf{Tx}$, $\mathsf{Total(player)}$ means the total number of the blockchain players, and $ \mathsf{\epsilon}$ denotes a secure threshold (e.g., 50\%) depending on a certain blockchain consensus algorithm.}
    
    \item[-] \textit{\textbf{Liveness.} As long as a transaction comes from an honest account holder, it will be rejected (a transaction does not store more than $k$ blocks deep) within time-bound $\delta$ by the honest blockchain players with a negligible probability, namely, $\mathsf{adv_{\adv,(k,\delta)}^{liveness}} =$}
    \begin{center}
    \begin{align*}
   \mathsf{ Pr}\left[
    \begin{array}{ll}
    \mathsf{while(t \leq \delta)\{}\\
    \qquad \mathsf{List(player) \gets FindPlayer(Tx,k,t);}\\
     \mathsf{\}}\\
     \mathsf{Num(List(player))/Total(player) \leq  \mathsf{\epsilon} }
   \end{array}
   \right] \leq \mathsf{negl(\lambda)},
   \end{align*}
\end{center}
    
\textit{where $\mathsf{Num(List(Tx))}$ represents the number of the blockchain players who have accepted $\mathsf{Tx}$, $\mathsf{Total(player)}$ means the total number of the blockchain players, and $ \mathsf{\epsilon}$ denotes a secure threshold (e.g., 50\%) depending on a certain blockchain consensus algorithm.}
\end{itemize}

\noindent Briefly speaking, the \textit{persistence} assumption says a transaction that has accepted by an honest player will be accepted (ended up at a depth of more than $k$ blocks) in other honest players' local chains. Meanwhile, the \textit{liveness} assumption states that all honest players will eventually agree on a decision or a value. The ``eventually" indicates that it may take a delay time $\delta$ ($\delta$ is finite) for reaching the agreement. By combining persistence and liveness, it ensures that the public ledger can only accept authentic transactions and will make them permanent.

\subsection{Smart Contract Formalization}
 \label{subsec-scprotocol}

From a high-level perspective, smart contracts are based on the form of state-machine replication~\cite{bessani2014state}. Thus, we simulate the smart contract as a distributed state machine. The states of contracts are replicated across different players in a distributed environment. The players participating in the system will automatically replicate the current state and transfer to a new state after a consensus round. In this procedure, blockchain systems act as virtual machines to provide an execution environment. 

\begin{defi}[$\widehat{\mathcal{SC}}$] \label{SC}
Smart contract is represented as a state machine by a tuple $\langle\mathsf{S},\mathsf{s}',\mathcal{T},f,\mathbb{B}  \rangle $, which is defined as: 
  \[ f:   \mathcal{S'} \xleftarrow{\mathbb{B}} \mathsf{S} \otimes  \mathcal{T}   ,\]
where  $\mathsf{S}$ represents a set of states or views, $\mathcal{S'}$ is the new state set after specified operations, $\mathcal{T}$ means the transactions that can trigger the execution of contract, $f$ is the transition function describing state changes.
\end{defi}

A complete execution of a smart contract in  blockchain systems consists of three procedures: \textit{contract deploy}, \textit{state transfer}, and \textit{state access}. The predefined logic can be coded into a file $\mathsf{bytecode}$ for the further deployment. Three sub-procedures are presented as follows.

\begin{itemize}
    \item[-] \textit{\textbf{Deploy}} $(\mathsf{\langle opcode  \rangle }, \mathsf{\langle reqcode  \rangle }, \mathsf{s}) \gets \mathsf{\langle bytecode  \rangle } \otimes \mathsf{Tx}.$ The deployment is triggered by a transaction $\mathsf{Tx}$ where $\mathsf{Tx}\in \mathcal{T}$. It takes as input the binary code $\mathsf{\langle bytecode  \rangle }$, and outputs initial state $\mathsf{s}$, where $\mathsf{s}\in \mathsf{S}$. The contract is compiled into $\mathsf{\langle opcode  \rangle }$ and $\mathsf{\langle reqcode  \rangle }$, where $\mathsf{\langle opcode  \rangle }$ specifies the operation set to be executed and $\mathsf{\langle reqcode  \rangle }$ defines the conditions depending on which the operation of $\mathsf{\langle opcode  \rangle }$ can be conducted.
    
    \item[-] \textit{\textbf{Transfer}} 
    $  \mathsf{s'} \xleftarrow{\mathbb{B}}  \mathsf{\langle input  \rangle } \otimes \mathsf{s} \otimes \mathsf{Tx'}. $ 
    By sending a transaction $\mathsf{Tx'}$ with an input $\mathsf{\langle input  \rangle }$, the current state $\mathsf{s}$ is transited to a new state $\mathsf{s'}$ under the operations on the blockchain system $\mathbb{B}$.

    \item[-] \textit{\textbf{Access}} $\mathsf{s'} \xleftarrow{\mathbb{B}} \mathsf{s'} \otimes \mathsf{Tx''}.$ By sending a query transaction $\mathsf{Tx''}$ through the blockchain $\mathbb{B}$, the state $\mathsf{s'}$ is returned by scanning the blockchain storage.

\end{itemize}

All the state information and instruction code are completely transparent. Any state and its changes are publicly accessible and publicly verifiable: (1) All users' transaction data and contract variables are visible to any observer; (2) The state change in a blockchain node will be verified by other nodes. 

\section{Generic Construction}

\label{sec-secureprotocol}

In this section, we provide a universal framework for smart contract-based security protocols. Then, we formalize their corresponding security properties: \textit{non-equivocation}, \textit{non-repudiation} and \textit{non-frameability}.

\subsection{Syntax of Contract-based Security Protocol}

Smart contract-based security protocols, noted as $\mathsf{\Pi}$, consist of two main types of roles: \textit{smart contract} and \textit{protocol users}. The smart contract is used to support a TTP for maintaining the information for encryption/decryption. The users are composed of both the message sender and the message receiver. Briefly speaking, the workflow is shown as follows. A message sender encrypts the message using the receiver's identity or the key under the assistance of a smart contract. This assistance is represented as storing or changing the newest state in the contract by sending a transaction. Afterwards, the message sender sends the ciphertext to the receiver. Then, the receiver decrypts the ciphertext by using the private key and public parameters fetched from the smart contract. Here, we emphasize the importance of the transactions that trigger the execution of a contract. Such transactions can be used as evidence to indicate the users' misbehaviours. A general construction is summarized as follows. Steps in black text run on the local client, while blue-text steps are executed on-chain. 

\smallskip
\hangindent=1em\noindent\textbf{System Setup} $\mathsf{pms} \gets \mathsf{Setup(1^\lambda)}$. The algorithm takes as input a security parameter $\lambda$, and outputs system parameters $\mathsf{pms}$.

\hangindent=1em\noindent\textbf{Key Generation} $\mathsf{(sk,pk)} \gets \mathsf{KeyGen(pms)}.$ The algorithm takes as input $\mathsf{pms}$, and outputs the receiver's key pair $\mathsf{(sk,pk)}$.

\smallskip
\noindent Then, a smart contract is deployed, with outputting a contract identity $\widehat{c}$, an initial state $\mathsf{\overline{s}}$, the operational code $\overline{\mathsf{opcode}}$, and the execution condition $\overline{\mathsf{reqcode}}$. The logic of a TPP is coded into $\overline{\mathsf{opcode}}$, and the execution condition of the logic is coded into $\overline{\mathsf{reqcode}}$. This step is finished by calling $\widehat{\mathcal{SC}}.\mathsf{Deploy}$ in Definition~\ref{SC}. Next, a message sender encrypts a message using the receiver's identity/key and auxiliary data with the assistance of the deployed smart contract. This assistance is represented as storing or changing auxiliary data in the contract by ways of sending transactions.

\hangindent=1em\noindent\textbf{\textcolor{blue}{Transaction Generation}} $\mathsf{Tx} \gets \mathsf{Sign(\sk_{tx},\mathsf{metadata}, aux)}.$ A user signs a transaction $\mathsf{metadata}$ with his private signing key $\sk_{tx}$ to obtain a transaction $\mathsf{Tx}$.

\hangindent=1em\noindent\textbf{\textcolor{blue}{ OnChain Operation}} $\mathsf{s} \xleftarrow{\mathbb{B}} \mathsf{ChainOpt}(\widehat{c}, \mathsf{\overline{s}},\mathsf{Tx}).$ The algorithm takes as input $\widehat{c}$, current state $\mathsf{\overline{s}}$ and a transaction $\mathsf{Tx}$ with auxiliary data $\mathsf{aux}$ used in previous steps, and outputs the transferred state $\mathsf{s}$ and the confirmed $\mathsf{aux}$. This algorithm is finished by calling the algorithm \textbf{State Transfer} described in Definition~\ref{SC}.

\hangindent=1em\noindent\textbf{Encryption} $ \mathsf{ct} \xleftarrow{} \mathsf{Enc}(\mathsf{\mathsf{pk},aux, m)}.$ The algorithm takes as input $\mathsf{pk}$, auxiliary data $\mathsf{aux}$ and a message $\mathsf{m}$, and outputs a cyphertext $\mathsf{ct}$. This algorithm is completed in the local client of users.


\smallskip
\hangindent=1em\noindent\textbf{\textcolor{blue}{State Read}} $\mathsf{s}^{\prime} \xleftarrow{\mathbb{B}} \mathsf{Read}(\widehat{c},\mathsf{Tx})$. The algorithm takes as input a contract identity $\widehat{c}$ and the transaction $\mathsf{Tx}$, and outputs a new state $\mathsf{s}^{\prime}$. 

\hangindent=1em\noindent\textbf{Decryption} $ \mathsf{m/\bot} \gets \mathsf{Dec}(\mathsf{\mathsf{sk}, s^{\prime}, ct})$. The algorithm takes as input $\mathsf{sk}$, $\mathsf{s}^{\prime}$, $\mathsf{ct}$, and outputs a message $\mathsf{m}$ or the special symbol $\bot$ indicating decryption failure. This algorithm is completed in the local client of users.

\hangindent=1em\noindent\textbf{\textcolor{blue}{Inspection}} 
$\mathsf{true/false} \xleftarrow{\mathbb{B}} \mathsf{Inspect}(\mathsf{Tx})$. This algorithm takes as input $\mathsf{Tx}$, and returns the legality of the \textbf{Transfer} operation.

\smallskip
\noindent The transaction $\mathsf{Tx}$ that triggers the execution of a contract in the \textbf{Transfer} operation is used as evidence to indicate the users' or TTP's misbehaviours, which significantly reduces the probability of committing malicious behaviours. 

\subsection{Oracles for Our Security Definition} 
To capture the security properties, a list of oracles modelling the honest parties are required. This section defines two oracles ($\mathsf{O}^{\mathsf{blockchain}}, \mathsf{O}^{\mathsf{user}}$) to simulate blockchain and an honest user. We use $\mathsf{(instruction; parameter)}$ to denote the instructions and inputs of oracles. Also, we define a list of (initially empty) sets $\mathcal{L}_1$, $\mathcal{L}_2$ and $\mathsf{Set(Tx)}$ to capture the output returned from oracles.

\smallskip
\noindent\textbf{Blockchain Oracle $\mathsf{O}^{\mathsf{bc}}:$} This oracle gives an adversary access to blockchain services. An adversary $\mathcal{A}$ can obtain the confirmed transaction and state. It provides the following interfaces. Here, the term ``confirmed'' means that the input date has been accepted by a blockchain system.

\begin{itemize}

\item[-] On input $(\mathsf{ReadState}; \mathsf{Tx})$, the oracle checks whether a tuple $(\mathsf{Tx},\mathsf{state}) \in \mathcal{L}_1$ exists, where $\mathsf{Tx}$ is an input for querying the state. If successful, the oracle returns $\mathsf{state}$ to $\mathcal{A}$; otherwise, it computes $\mathsf{state} \gets \mathbb{B}.\mathsf{ReadState(Tx)}$ and adds $(\mathsf{Tx},\mathsf{state})$ to $\mathcal{L}_1$, and then returns $\mathsf{state}$ to $\mathcal{A}$. 

\item[-] On input $(\mathsf{ReadTx}; \mathsf{state})$, the oracle checks whether a tuple $(\mathsf{state},\mathsf{Tx}) \in \mathcal{L}_2$ exists, where $\mathsf{state}$ is a confirmed state for querying the transaction that triggers the execution of
state updating. If successful, the oracle returns $\mathsf{Tx}$ to $\mathcal{A}$; otherwise, it computes $\mathsf{Tx} \gets \mathbb{B}.\mathsf{ReadTx(Tx)}$ and adds $(\mathsf{state},\mathsf{Tx})$ to $\mathcal{L}_2$, and then returns $\mathsf{Tx}$ to $\mathcal{A}$. 
\end{itemize}


\smallskip
\noindent\textbf{User Oracle $\mathsf{O}^{\mathsf{user}}:$} This oracle simulates an honest user. It gives an adversary access to valid transactions, and provides the following interfaces.

\begin{itemize}

\item[-] On input $( \mathsf{Sign\;}; \mathsf{metadata})$, the oracle $\mathsf{O}^{\mathsf{user}}$ checks whether $(\mathsf{metadata},\mathsf{Tx}) \in \mathsf{Set(Tx)}$ exists, where $\mathsf{metadata}$ is an input of a transaction. If it is successful, the oracle returns $\mathsf{Tx}$ to $\mathcal{A}$; otherwise, it computes $\mathsf{Tx} \gets \mathsf{Sign(\cdot,\mathsf{metadata, aux})}$ and adds $(\mathsf{metadata},\mathsf{Tx})$ to $\mathsf{Set(Tx)}$, and then returns $\mathsf{Tx}$ to $\mathcal{A}$. 

\end{itemize}
\subsection{Security Definition}
In this section, we provide three well-defined properties that relate to the protocol security. These properties show how smart contracts strengthen their applied protocols.


\smallskip
\noindent\textbf{Non-equivocation.} Non-equivocation ensures that a smart contract shares the same state for encryption/decryption. Alternatively, after a transaction's invocation, a unique and deterministic state should be obtained, and the smart contract should share the same data view to both the message sender and message receiver. To formalize this property, we consider a game in which the adversary is allowed to interact with the blockchain oracle $\mathsf{O}^{\mathsf{blockchain}}$. Assume that $\mathsf{\mathsf{Tx}}$ is a transaction submitted to call the smart contract from an adversary $\mathcal{A}$ and $\mathsf{s}$ is the corresponding execution result. The adversary wins if he gets another valid state $\mathsf{s^{\star}}$ by querying $\mathsf{O}^{\mathsf{blockchain}}$, where $\mathsf{s^{\star}}$ does not equal to $\mathsf{s}$ and $\mathsf{s^{\star}}$ is not in $\mathcal{L}_1$, but the transaction $\mathsf{Tx}$ is indeed stored in the blockchain. We define $\mathsf{adv_{\adv,{\mathsf{\Pi}}}^{\Game_{neqv}}(\lambda)} = \mathsf{Pr[G_{\adv,{\mathsf{\Pi}}}^{neqv}(\lambda)}]$, where $\mathsf{G_{\adv,\mathsf{\Pi}}^{neqv}(\lambda)}$ is defined as follows.

\begin{center}
\fbox{
 \resizebox{0.55\linewidth}{!}{ 
\procedure[linenumbering]{$\mathsf{G}_ {\mathcal{A},\mathsf{\Pi}}^{\textrm{neqv}}(\lambda)$}{ %
\mathsf{pms} \gets \mathsf{Setup(1^\lambda)} \\
(\mathsf{pk},\mathsf{sk}) \gets \mathsf{KeyGen(pms)}. \\
\mathsf{Tx \gets Sign(sk_{tx}, metadata, aux)}  \\ 
\mathsf{s} \xleftarrow{\mathbb{B}} \mathsf{ChainOpt}(\widehat{c}, \mathsf{\overline{s}},\mathsf{Tx}) \\
\mathsf{ct} \xleftarrow{\;} \mathsf{Enc}(\mathsf{\mathsf{pk}, aux, m)} \\
\mathsf{s^{\star}} \xleftarrow{\text{query}}  
\adv^{\mathsf{O}^{\mathsf{bc(ReadState;Tx)}}}(\widehat{c},...)
\\ 
\mathsf{m} \gets \mathsf{Dec}(\mathsf{\mathsf{sk}, \mathsf{s^{\star}}, ct}) \\
\pcreturn (\mathsf{s^{\star}} \notin \mathcal{L}_1) \wedge (\mathsf{s^{\star}} \neq \mathsf{s}) \wedge (\mathsf{true} = \mathsf{Inspect}(\mathsf{Tx}))
}}  
}
\end{center}

\begin{defi}
A hybrid protocol $\mathsf{\mathsf{\Pi}}$ is said to achieve the property of non-equivocation, if for all probabilistic polynomial time (PPT) adversaries $\mathcal{A}$, there exists a negligible function $ \mathsf{negl(\lambda)}$ satisfying   $\mathsf{adv_{\adv,{\mathsf{\Pi}}}^{\Game_{neqv}}(\lambda) < negl(\lambda)}$.
\end{defi}

\smallskip
\noindent\textbf{Non-repudiation.} Non-repudiation is, fundamentally, the requirement that a user
cannot deny having executed a certain function in a smart contract. This indirectly indicates that the transaction triggered execution of smart contracts, cannot be tampered with. To formalize the \textit{non-repudiation}, we consider a game in which an adversary $\mathcal{A}$ interacts with a contract. Assume that $\mathsf{\mathsf{Tx}}$ is a transaction submitted to blockchain from adversaries and $\mathsf{s}$ is the corresponding answer. The adversary wins if he gets another valid transaction and state pair $(\mathsf{Tx^{\star}}, \mathsf{s}^{\star})$, where $\mathsf{s}$ equals to $\mathsf{s}^{\star}$ and $\mathsf{Tx^{\star}}$ does not match $\mathsf{Tx}$. We define  $\mathsf{adv_{\adv,\mathsf{\Pi}}^{\Game_{nrep}}(\lambda)} = \mathsf{Pr[G_{\adv,\mathsf{\Pi}}^{nrep}(\lambda)}]$, where $\mathsf{G_{\adv,\mathsf{\Pi}}^{nrep}(\lambda)}$ is defined as follows.

\begin{center}
\fbox{
 \resizebox{0.55\linewidth}{!}{ 
\procedure[linenumbering]{$\mathsf{G}_ {\mathcal{A},\mathsf{\Pi}}^{\textrm{nrep}}(\lambda)$}{ %
\mathsf{pms} \gets \mathsf{Setup(1^\lambda)} \\
(\mathsf{pk},\mathsf{sk}) \gets \mathsf{KeyGen(pms)}. \\
\mathsf{Tx \gets Sign(sk_{tx}, metadata, aux)}  \\ 
\mathsf{s} \xleftarrow{\mathbb{B}} \mathsf{ChainOpt}(\widehat{c}, \mathsf{\overline{s}},\mathsf{Tx}) \\
\mathsf{ct} \xleftarrow{\;} \mathsf{Enc}(\mathsf{\mathsf{pk}, aux, m)} \\
\mathsf{s}^{\prime} \xleftarrow{\mathbb{B}} \mathsf{Read}(\widehat{c},\mathsf{Tx}) \\
\mathsf{Tx^{\star}} \xleftarrow{\text{query}}  
\adv^{\mathsf{O}^{\mathsf{blockchain(ReadTx;^{\prime})}}}(\widehat{c},...)
\\
\mathsf{m} \gets \mathsf{Dec}(\mathsf{\mathsf{sk}, \mathsf{s}, ct}) \\
\pcreturn (\mathsf{Tx^{\star}} \notin \mathcal{L}_2) \wedge (\mathsf{Tx^{\star}} \neq \mathsf{Tx}) \wedge (\mathsf{true} = \mathsf{Inspect}(\mathsf{Tx^{\star}}))
}}  
}
\end{center}

\begin{defi}
A hybrid protocol $\mathsf{\mathsf{\Pi}}$ is said to achieve the property of non-repudiation, if for all PPT adversaries $\mathcal{A}$, there exists a negligible function $ \mathsf{negl(\lambda)}$ satisfying $\mathsf{adv_{\adv,\mathsf{\Pi}}^{\Game_{nrep}}(\lambda) < negl(\lambda)}$.
\end{defi}

\smallskip
\noindent\textbf{Non-frameability.} The non-frameability property indicates that a user cannot be framed by producing evidence of his ``misbehaviour''. As discussed before, to trigger a contract execution, a transaction is required to sent from a user. If an honest user does not invoke the function in a contract, he will never be wrongfully accused. Formally, we assume that $\mathsf{Set(\mathsf{Tx})}$ is a collection of queries submitted to the user oracle from adversaries. The adversary wins if he successfully imitates an honest user with the transaction $\mathsf{Tx^{\star}}$ and achieves a valid state $\mathsf{s^{\star}}$ for decryption. We define  $\mathsf{adv_{\adv,\mathsf{\Pi}}^{\Game_{nfrm}}(\lambda)} = \mathsf{Pr[G_{\adv,\mathsf{\Pi}}^{nfrm}(\lambda)}]$, where $\mathsf{G_{\adv}^{nfrm}(\lambda)}$ is defined as,

\begin{center}
\fbox{
 \resizebox{0.55\linewidth}{!}{ 
\procedure[linenumbering]{$\mathsf{G}_ {\mathcal{A},\mathsf{\Pi}}^{\textrm{nfrm}}(\lambda)$}{ %
\mathsf{pms} \gets \mathsf{Setup(1^\lambda)} \\
(\mathsf{pk},\mathsf{sk}) \gets \mathsf{KeyGen(pms)}. \\
\mathsf{Tx^{\star}} \xleftarrow{\text{query n times}}  
\adv^{\mathsf{O}^{\mathsf{user}}(Sign;metadata)}(\mathsf{metadata})
\\ 
\mathsf{s} \xleftarrow{\mathbb{B}} \mathsf{ChainOpt}(\widehat{c}, \mathsf{\overline{s}},\mathsf{Tx}) \\
\mathsf{ct} \xleftarrow{\;} \mathsf{Enc}(\mathsf{\mathsf{pk}, aux, m)} \\
\mathsf{s}^{\prime} \xleftarrow{\mathbb{B}} \mathsf{Read}(\widehat{c},\mathsf{Tx}) \\
\mathsf{m} \gets \mathsf{Dec}(\mathsf{\mathsf{sk}, \mathsf{s}^{\prime}, ct}) \\
\pcreturn (\mathsf{Tx^{\star}} \notin \mathsf{Set(Tx)}) \wedge (\mathsf{true} = \mathsf{Inspect}(\mathsf{Tx^{\star}}))
}}  
}
\end{center}

\begin{defi}
A hybrid protocol $\mathsf{\mathsf{\Pi}}$ is said to achieve the property of non-frameability, if for all PPT adversaries $\mathcal{A}$, there exists a negligible function $\mathsf{negl(\lambda)}$ satisfying  $\mathsf{adv_{\adv,{\mathsf{\Pi}}}^{\Game_{nfrm}}(\lambda) < \mathsf{negl(\lambda)}}$.\\
\end{defi}

\section{Instantiations}
\label{sec-instance}

In this section, we provide two concrete constructions by using our formalized SC framework. The first instance is to enable certificate revocation transparent in CBE, while the second instance is to establish a transparent RBE scheme, making Key Curator's operations publicly visible and checkable.

\begin{figure*}[!htb]
    \centering
    \includegraphics[width=\linewidth]{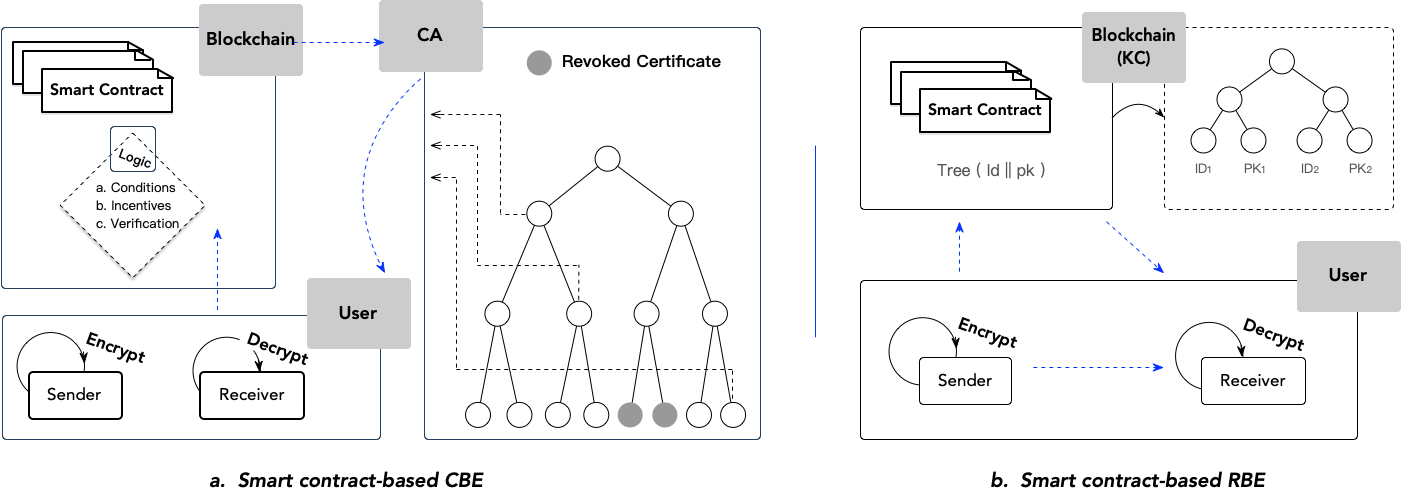}
   \caption{Smart contract-based security protocols}
    \label{fig:construction}
\end{figure*}

\subsection{Making Revocation Transparent for CBE}

In this case, we introduce a transparent certificate revocation mechanism for CBE (cf. Fig.~\ref{fig:construction}-a). The smart contract acts as an agent to assist the Certificate Authority(CA) in managing the revocation. Specifically, users are required to send revocation requests to a smart contract. This is achieved by sending a transaction that contains the revocation information. Then, the smart contract checks the validity of the requests, including the authenticity of identity, the expiry date of the certificate, etc. Next, the smart contract periodically transfers the valid requests to CA. Finally, CA releases the new \textit{reconfirmation} status (stopping the issuance of certificates for the revoked public key). The hybrid protocols are as follows.

\smallskip
\hangindent=1em\noindent\textbf{System Setup.}$\mathsf{pms} \gets \mathsf{Setup(1^\lambda)}$. The algorithm takes as input a security parameter $\lambda$, (optionally) the total number of time periods $\mathsf{n}$, and outputs $\mathsf{pms}$. $\mathbb{G}_1$ and $\mathbb{G}_2$ are two cyclic groups of some large prime order $q$.

\begin{align*}
   \hat{e}: \mathbb{G}_1 \times \mathbb{G}_1 \gets \mathbb{G}_2 \\
   P \in \mathbb{G}_1 \\
   H_1 : \{0, 1\}^{\star} \to G_1, \quad H_2 :  \mathbb{G}_1 \to \{0, 1\}^n \\
   s_C \in \mathbb{Z}/q\mathbb{Z} \\
   Q = s_C P \\
   \mathsf{pms} = (\mathbb{G}_1, \mathbb{G}_2, \hat{e}, P, Q, H_1, H_2) \\
\end{align*}

The system parameters are $\mathsf{pms}$. The message
space is $\mathcal{M} = \{0,1\}^n$. The CA’s secret key is $s_C$. 

\smallskip
\hangindent=1em\noindent\textbf{Key Generation} $(s_B,p_B) \gets \mathsf{KeyGen(pms)}$. The algorithm takes as input $\mathsf{pms}$, and outputs a user's (e.g., Bob's) key pair $(s_B,p_B)$ for encryption/decryption.

\begin{align*}
   s_B \in \mathbb{Z}/q\mathbb{Z} \\
   p_B = s_BP
\end{align*}

\noindent This algorithm is run by users. $s_B$ is a user's (Bob's) private key, which is a random number selected in $\mathbb{Z}/q\mathbb{Z}$ and $s_BP$ is computed according to the parameters issued by the CA.

\smallskip
\hangindent=1em\noindent\textbf{\textcolor{blue}{Transaction Generation}} $\mathsf{Tx} \gets \mathsf{Sign(\sk_{tx},\mathsf{metadata}, aux)}.$ The algorithm signs a transaction $\mathsf{metadata}$ with a private signing key $\sk_{tx}$ to obtain a transaction $\mathsf{Tx}$. Here, $\mathsf{aux}$ represents a request mapping to certificate state. For example, $\mathsf{aux} = [bob:revoked]$. This algorithm is run by a user who wants to revoke his certificate.

\smallskip
\hangindent=1em\noindent\textbf{\textcolor{blue}{OnChain Operation}}. It consists of the following sub-algorithms.

\begin{itemize}
    
    \item[-] $\textit{Contract Deploy}$  $\widehat{c}, \mathsf{\overline{s}} \xleftarrow{\mathbb{B}}  \widehat{\mathcal{SC}}.\mathsf{Deploy(bytecode)}$. CA runs the algorithm by inputting $bytecode$, and outputs a contract $\widehat{c}$, an initial $\mathsf{\overline{s}}$ and operational code $\overline{\mathsf{opcode}}$, and execution conditions $\overline{\mathsf{reqcode}}$.

    \begin{center}

\fbox{
\parbox{0.95\textwidth}{%
\smallskip

\centerline{Functionalities defined in the contract $\widehat{c}$.}
\smallskip

 $\overline{\mathsf{reqcode}}: \textbf{Revocation Qualification Check}$. 
 Once received the certificate revocation request, the contract $\widehat{c}$ checks the revocation qualification based on policies predefined in $\overline{\mathsf{reqcode}}$.

 \smallskip
 $\overline{\mathsf{opcode}}: \textbf{Certificate Data Update}$. stores new request to a public list and informs the CA that a user's revocation requests are ready.
}%
}
\end{center}


  \item[-] $\textit{State Transfer}$ $\mathsf{s} \xleftarrow{\mathbb{B}} \widehat{\mathcal{SC}}.\mathsf{Transfer} (\widehat{c}, \mathsf{\overline{s}},\mathsf{Tx})$ The algorithm takes as input a contract identity $\widehat{c}$, a contract's initial state $\overline{\mathsf{s}}$, $\mathsf{Tx}$, and outputs the transferred state $\mathsf{s}$. Here, $\mathsf{s}$ refers to as users' certificate state (see Table~\ref{tab:cert_state}).

\end{itemize}


\begin{table*}[!hbt]
 \caption{on chain state of user's certificate state}\label{tab:cert_state}
 \label{node}
  \centering
    \resizebox{0.6\linewidth}{!}{  
    \begin{tabular}[t]{cccc}
    \toprule
     \quad \textbf{Number}\quad   & \quad\textbf{User ID}\quad   & \quad \textbf{State}\quad  & \quad \textbf{Expiry Date}\quad  \\  \midrule
     1 & Alice & valid  & Dec, 2022 \\  \midrule 
     2 &  Bob  & revoked  & Dec, 2021 \\   
     3 & Tom & revoked & Jan, 2022\\ \midrule
     4 & Kate & valid & Jun, 2022\\
     5 & David & valid & Nov, 2022\\
    \bottomrule
    \end{tabular}
    }
\end{table*}

\noindent The CA needs to fetch the latest state from the contract. Thus, it covers two sub-algorithms.

\smallskip
\hangindent=1em\noindent\textbf{\textcolor{blue}{State Read}}  $\mathsf{s}^{\prime} \xleftarrow{\mathbb{B}} \mathsf{Read}(\widehat{c},\mathsf{Tx})$. CA runs this algorithm by inputting a contract $\widehat{c}$, and outputs the confirmed state $\mathsf{s}^{\prime}$ containing the user's valid revocation requests (Bob and Tom's revocation state, as shown in Table~\ref{tab:cert_state}). 
  
\smallskip
\hangindent=1em\noindent\textbf{Certificate} $\mathsf{  \mathsf{Cert_{i}} \gets Cert( msk,i,user,pk, s^{\prime}) }$. 
  CA takes as input a certifier’s master secret $\mathsf{msk}$, user's information $\mathsf{user}$, public key $\mathsf{pk}$ and the updated state $\mathsf{s^{\prime}}$, and outputs the certificate $\mathsf{Cert_{i}}$. To be specific, CA updates certificates' status through a binary tree. CA arranges for most $2^m$ clients as leaves in a $m$-level binary tree. Each client is embedded by a unique $m$-bit serial number (SN) in its leaf nodes, and SN provides both identities and positions in the tree. The revocation is represented by the deletion of a leaf's sub-cover nodes. Meanwhile, to improve the efficiency of updating, the difference sub-cover approach~\cite{naor2001revocation} can be adopted. Note that $S_{ij}$ denotes the set of leaves in the subset of $S_i$ but not in $S_j$. The contract state $\mathsf{s^{\prime}}$ records the latest request from users. Only eligible revocation requests (e.g., valid time or matched identities) can be approved for this state.

  \begin{figure*}[!htb]
    \centering
    \includegraphics[width=0.6\linewidth]{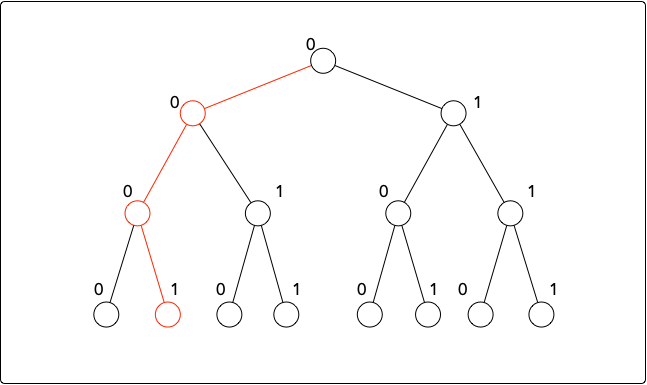}
   \caption{$m$-level binary tree by embedding a unique $m$-bit serial number}
    \label{fig:mapping}
\end{figure*}

\begin{align*}
H_5: \{0, 1\}^\star \to \mathbb{G}_1 \\
x \in \mathbb{Z}/q\mathbb{Z} \\
T_i = H_5(Q, i) \\
P_k = H_1(b_1 \dots b_k) \\
Cert_{i} = s_C T_i + xP_k \\
\end{align*}

\smallskip
\hangindent=1em\noindent\textbf{Encryption} $\mathsf{ct} \gets \mathsf{Enc}(\mathsf{Q,BobInfo, m)}$. At the time period $\mathsf{i}$, the algorithm takes as input CA's public key $\mathsf{Q}$, a message $\mathsf{m}$, and outputs the ciphertext $\mathsf{ct}$.
In this algorithm, $r$ is random number, $H_5$ maps $\{0,1\}^\star$ to $\mathbb{G}_1$ points.

\begin{align*}
m \in \mathcal{M} = \{0,1\}^{n} \\
r \in \mathbb{Z}/q\mathbb{Z}  \\
P_B' = H_1(\mathsf{BobInfo}) \in G_1 \\
T_i = H_5(\mathsf{Q}, i)\\
g = \hat{e}(\mathsf{Q}, T_i)\hat{e}(s_BP, P_B') \\
V = M \otimes H_2(g^r) \\
\mathsf{ct} = [rP, rP_1, . . . , rP_m, V ] \\
\end{align*}

\noindent Note that, at this stage, the message sender has already verified Bob’s initial certificate, and therefore knows $\mathsf{BobInfo}$.

\smallskip
\hangindent=1em\noindent\textbf{Decryption} $  \mathsf{m/\bot \gets Decrypt}(\mathsf{s_B,  \mathsf{Cert_{i}}, ct})$. At the time period $\mathsf{i}$, the algorithm takes as input a secret key $s_B$, a reconfirmation certificate $ \mathsf{Cert_{i}}$ (if it exists), and outputs a message $\mathsf{m}$ or a special symbol $\bot$ indicating decryption failure.

\begin{align*}
   \mathsf{Cert_{i}} = s_C T_i + xP_k \\
   M = V \otimes H_2(\frac{\hat{e}(rP,S_i+s_BP_B')}{\hat{e}(xP,rPk)}).
\end{align*}

\hangindent=1em\noindent\textbf{\textcolor{blue}{Inspection}} 
$\mathsf{true/false} \xleftarrow{\mathbb{B}} \mathsf{Inspect}(\mathsf{Tx})$. This algorithm takes as input $\mathsf{Tx}$, and returns the legality of the \textbf{Transfer} operation. $\mathsf{true}$ indicates that the revoked certificate is under the users' intention.

\subsection{Building Transparent KC for RBE}

This instance builds a transparent RBE by using smart contract (see Fig.~\ref{fig:construction}-b). A smart contract is employed as a Key Curator (KC) to maintain the relationship between the identity and the corresponding public key, and based on that, and it returns some public parameters as the encryption key. The \textit{users} are composed of the role of the message sender and the message receiver. A high-level workflow is shown as follows. A message receiver registers his identity and key binding to a smart contract-based KC by sending a transaction. After the blockchain has confirmed the transaction that containing this identity-key binding, a public parameter is obtained. Now, a message sender is allowed to encrypt a message using the receiver's identity and public parameters, and then the sender sends this ciphertext to the receiver. Afterwards, the receiver decrypts the ciphertext using the private key and the public parameter updated from the smart contract. Here, we emphasize that all the KC's operations, including user's registration and identity-key pair compression, are executed on-chain, which is publicly auditable.

\smallskip
\hangindent=1em\noindent{\textbf{System Setup}} $\mathsf{pms} \gets \mathsf{Setup(1^\lambda)}$. The algorithm takes as input a security parameter $\lambda$ and outputs a common random string $\mathsf{crs}$, an empty auxiliary information $\mathsf{aux_{tree} = \varnothing}$ and public parameters $\mathsf{pp_0 = (hk_1,\dots,hk_{\lambda})}$, where each $\mathsf{hk_i}$ is sampled from $\mathsf{HGen(1}^\lambda, \mathsf{0)}$ (cf. Appendix A). 


\smallskip
\hangindent=1em\noindent{\textbf{Key Generation}} $\mathsf{(pk,sk)} \gets \mathsf{KeyGen(pms)}$. This algorithm takes as input  $\mathsf{pms}$, and outputs a public key $\mathsf{pk}$ and a secret key $\mathsf{sk}$. The algorithm is run by any honest party who wants to register his identity to the RBE system. The secret key $\mathsf{sk}$ is privately held by a user, while $\mathsf{pk}$ is freely accessible by the public.

\smallskip
\hangindent=1em\noindent\textbf{\textcolor{blue}{Transaction Generation}} $\mathsf{Tx} \gets \mathsf{Sign(\sk_{tx},\mathsf{metadata}, aux)}.$ A user signs a transaction $\mathsf{metadata}$ with his private signing key $\sk$ to obtain a transaction $\mathsf{Tx}$. This procedure is represented as a registration, and $\mathsf{aux}$ refers to the registration information including a user's identifier $\mathsf{id}$ and public key $\mathsf{pk}$.

\smallskip
\hangindent=1em\noindent{\textbf{\textcolor{blue}{OnChain Operation}}}. The algorithm takes as input the auxiliary information $\mathsf{\mathsf{s^\star} := (Tree_1, \dots, Tree_{\eta})}$ ($i$ is the index of the tree that holds $\mathsf{id}$) from internal states of the contract, $\mathsf{id}$ from users, and outputs a Merkle opening of the path that leads to $\mathsf{id}$ to the root in $\mathsf{Tree_i}$. The path means the set of sibling nodes count from a leaf to the root. We represent it as $\mathsf{aux := [(h_0^0, h_0^1), (h_1^0,  h_1^1,  b_1), \dots,}\\ \mathsf{(h_{d_i - 1}^0,  h_{d_i - 1}^1,  b_{d_i - 1}), rt_i]}$. $\mathsf{h_0^0}$ refers to a user's identifier and $\mathsf{h_0^1}$ refers to the user's public key. $\mathsf{rt_i}$ is the root of the tree and $\mathsf{d_i}$ is the depth of the tree, and $\mathsf{b_i} \in \mathsf{\{left,right\}} $. This algorithm consists of three sub-algorithms: identity verification, parameter parsing and identity-key pair compression. 



\smallskip
\hangindent=1em\noindent ${\textit{\textbf{Identity Verification}}}$. Once received the identity registration request, the contract $\widehat{c}$ checks the identity based on the pre-defined logic. The sub-algorithm prevents registering multiple keys for already registered users or registering any key for currently unregistered users. 
  

\hangindent=1em\noindent ${\textit{\textbf{Parameter Parsing}}}$. After that, the smart contract $\widehat{c}$ parses the parameter through retrieve internal state stored in $\widehat{c}$. Intuitively, when a new user joins in the system, the tree root is updated and the public parameters of registered users also need to be updated. To minimize the effect of registration by new users on previously registered users, our solution, following the idea of the original RBE \cite{RBE18}, adopts multiple Merkle hash trees such that any individual user is affected only a limited number of times. In particular, the trees with the same depth are continuously merged in a new one in the on-chain calculation (see Fig.\ref{fig:construction}); the tree that holds the identity only needs to be updated at most $\mathcal{O}(\mathsf{log n)}$ times, where $\mathsf{n}$ represents the total number of registered users. Thereby, a registered user does not have to query the smart contract  each time for public parameters.  

\begin{itemize}
\item Parse $\mathsf{aux_{tree}} := \mathsf{((Tree_1,\dots, Tree_{\eta}),(id_1, \dots, id_n))}$ where the trees have corresponding depths $ \mathsf{d_1 \textgreater d_2 \dots \textgreater d_{\eta}}$, and $\mathsf{(id_1,\dots, id_n)}$ is the order by which the current identities have registered. 

\item Parse $\mathsf{pp_n}$ as a sequence $\mathsf{((hk_1,\dots, hk_{\lambda}),(rt_1, d_1),\dots ,(rt_{\eta},d_{\eta}))}$, and $\mathsf{rt_i \in \{0, 1\}}^\lambda$ represents the root of $\mathsf{Tree_i}$ while $\mathsf{d_i}$ is the depth of $\mathsf{Tree_i}$.

\item Create the new tree $\mathsf{Tree_{\eta+1}}$ with leaves $\mathsf{id}$ and $\mathsf{pk}$. Then, set its root as $\mathsf{rt_{\eta+1}:= Hash(hk_1, id||pk)}$ and thus its depth would be $\mathsf{d_{\eta+1} = 1}$.
\end{itemize}

\hangindent=1em\noindent ${\textit{\textbf{Identity-key Pair Compression}}}$. Then, the smart contract $\widehat{c}$ starts to merge multiple Merkle hash trees through the sub-algorithm $\mathsf{SC.Transfer}$. Let $\mathcal{T} = \{\mathsf{Tree_1, . . . , Tree_{\eta+1}}\}$. While there are two different trees $\mathsf{Tree_L}, \mathsf{Tree_R} \in \mathcal{T}$ of the same depth $\mathsf{d}$, same size $\mathsf{s = 2^d}$ (as our trees are full binary trees), the algorithm keeps doing the following steps.

\begin{itemize}

\item Let $\mathsf{Tree}$ be a new tree of depth $\mathsf{d + 1}$ that contains $\mathsf{Tree_L}$ as its left subtree, $\mathsf{Tree_R}$ as right subtree, and $\mathsf{rt = Hash(hk_{d+1},rt_L||rt_R)}$ as the root.

\item Remove both of $\mathsf{Tree_L}$,$\mathsf{Tree_R}$ from $\mathcal{T}$ and add $\mathsf{Tree}$ to $\mathcal{T}$ instead.

\item Let $\mathcal{T} \mathsf{:=(Tree_1, \dots, Tree_{\zeta})}$ be the final set of trees with depths $\mathsf{d_1^{'} \textgreater \dots \textgreater d_{\zeta}^{'}}$ and roots $\mathsf{rt_1^{'},\dots, rt_{\zeta}^{'}}$. Set the $\mathsf{pp_{n+1}}$ and $\mathsf{aux_{tree}}$ as follows:

\begin{itemize}
\item $\mathsf{pp_{n+1} := ((hk_1, \dots, hk_{\lambda}),(rt_1^{'}, d_1^{'}),\dots,(rt_{\zeta}^{'},d_{\zeta}^{'}))}$, 
\item $ \mathsf{\mathsf{aux_{tree}} := (\mathcal{T} ,(id_1, \dots, id_n, id_{n+1} = id))}.$
\end{itemize}

\end{itemize}

\smallskip
\hangindent=1em\noindent\textbf{Encryption}
$\mathsf{ct} \gets \mathsf{Enc}(\mathsf{crs, pp;id, m)}$. The algorithm takes as input the common random string $\mathsf{crs}$, a public parameter $\mathsf{pp}$ that obtained from the contract, an identity $\mathsf{id}$, a message $m$, and outputs the ciphertext $\mathsf{ct}$. In particular, it parses $\mathsf{pp := ((hk_1, \dots, hk_{\lambda}),(rt_1, d_1), \dots ,}$ $\mathsf{(rt_{\eta}, d_{\eta}))}$. Then, it generates programs $\mathsf{P_1, \dots, P_\eta}$ where $\mathsf{P_i}$ works as follows,

\smallskip
\noindent\textit{Hardwired values}: $\mathsf{rt_i, d_i,(hk_1, \dots, hk_{d_i}), m, id, r }$ (randomness); Note that $\mathsf{hk_i}$ corresponds to the level $i$ in a Tree. \\
\smallskip
\noindent\textit{Input: $\mathsf{pth}$}
\begin{itemize}
\item  Parse $\mathsf{pth := [(h_0^0, h_0^1), (h_1^0,  h_1^1,  b_1), \dots, (h_{d_i - 1}^0,  h_{d_i - 1}^1, b_{d_i - 1}),  }$
$\mathsf{rt]}$

\item If $\mathsf{rt_i \neq rt}$, then output $\mathsf{\bot}$.

\item If $\mathsf{id \neq h_0^0}$, then output $\mathsf{\bot}$.

\item If $\mathsf{rt = Hash(hk_{d_i}, h_{d_i - 1}^0||h^1_{d_i - 1})}$ and $\mathsf{h^{b_j}_j = Hash(hk_j, h
^0_{j - 1} ||h^1_{j - 1})}$ for all $\mathsf{j \in [d_i - 1]}$, then output $\mathsf{\mathsf{PKE.Enc}(h^1_0, m;r)}$(see Appendix A) by using $\mathsf{h^1_0}$ as the public key and $\mathsf{r}$ as the randomness, otherwise output $\bot$.

\end{itemize}
Finally, the algorithm outputs $\mathsf{ct := (pp, Obf(P_1),\dots Obf(P_{\eta}))}$ where $\mathsf{Obf}$ specifies the IO obfuscation. To be specific, the encryption is performed by an obfuscation of the program $\mathsf{P}$ using the public key and some auxiliary information that connects the public key. In particular, The program $\mathsf{P}$ outputs an encryption of $m$ only if the path is a \textit{Merkle opening} for leaves $\mathsf{(id, pk)}$ within the Merkle tree with root $\mathsf{h}$. When there are multiple trees $\mathsf{Tree_1, \dots Tree_\eta}$ held by the smart contract, the ciphertext includes $\eta$ obfuscations,  one for every tree $\mathsf{Tree_i}$ $(i \leq \eta)$. 

\smallskip
\hangindent=1em\noindent\textbf{\textcolor{blue}{State Read}} $\mathsf{s}^{\prime} \xleftarrow{\mathbb{B}} \mathsf{Read}(\widehat{c},\mathsf{Tx})$. The algorithm takes as input a contract identity $\widehat{c}$ and the transaction $\mathsf{Tx}$, and outputs a new state $\mathsf{s}^{\prime}$.

\smallskip
\hangindent=1em\noindent\textbf{Decryption}
$ \mathsf{ m/\bot \gets Dec}(\mathsf{sk, \mathsf{s}^{\prime}, ct})$. The algorithm takes as input the secret key $\mathsf{sk}$, an updated Merkle opening $\mathsf{aux}$ extracted from $\mathsf{s}^{\prime}$, a ciphertext $\mathsf{ct}$, and outputs the message $\mathsf{m} \in \{0, 1\}^*$ or in $\{\perp, \mathsf{GetUpd}\}$. The special symbol $\perp$ indicates a syntax error, while $\mathsf{GetUpd}$ indicates that the latest received information needs to be updated by re-executing the algorithms of
\textit{State Read} and 
\textit{Operation}. In particular, it parses $\mathsf{ct = (u,\overline{P}_1, \dots, \overline{P}_{\eta})}$, and then executes $\mathsf{m_i = \mathsf{PKE.Dec}(sk,\overline{P}_i(u))}$ for each program $\mathsf{\overline{P}_i}$, and finally outputs the message satisfies $\mathsf{m_i \neq \{\perp, \mathsf{GetUpd}\}}$.

\hangindent=1em\noindent\textbf{\textcolor{blue}{Inspection}} 
$\mathsf{true/false} \xleftarrow{\mathbb{B}} \mathsf{Inspect}(\mathsf{Tx})$. This algorithm takes as input $\mathsf{Tx}$, and returns the legality of the \textbf{Transfer} operation.

\section{Security Proof}
\label{sec-proof}
In this section, we provide the security proofs of the concrete constructions in terms of aforementioned properties.

\begin{thm}[Non-equivocation]
Assume that the blockchain is robust, our construction satisfies the property of non-equivocation.
\label{non-equivocation}
\end{thm}


\begin{prf}
The non-equivocation indicates that two public states gained from the smart contracts under the same query condition (e.g., same $\overline{\mathsf{opcode}}$) are different. Assume that there is an adversary $\mathcal{A}$ that is able
to win the game $\mathsf{G_{\adv,\mathsf{\Pi}}^{neqv}(\lambda)}$. Then, we build an adversary $\mathcal{B}$ against the promise of persistence and liveness. Using the adversary $\mathcal{A}$'s advantage, $\mathcal{B}$ can obtain two different states $\mathsf{s}'$ and $\mathsf{s}$ under the invocation of a certain transaction $\mathsf{Tx}$. Without loss of the generality, we assume $\mathsf{s}$ is a valid state and $\mathsf{s}'$ is an invalid state. Also, for simplicity, we use $\mathsf{\mathcal{F}(Tx, state) = accepted}$ to represent that a state has been ``confirmed" by the major blockchain players in the time-bound $\delta$. Here, the ``confirmed" means the player has stored a transaction more than $k$ blocks deep. Then, four events (denoted by $\mathbb{E}$) may happen when $\mathcal{B}$ has fetched two different states.

\begin{numcases}{}
\mathsf{\mathcal{F}(Tx, s) = accepted} \wedge \mathsf{\mathcal{F}(Tx, s^{\star}) \neq accepted} \\ 
\mathsf{\mathcal{F}(Tx, s^{\star}) = accepted} \wedge \mathsf{\mathcal{F}(Tx, s) \neq accepted} \\ 
\mathsf{\mathcal{F}(Tx, s^{\star}) \neq accepted} \wedge \mathsf{\mathcal{F}(Tx, s) \neq accepted} \\ 
\mathsf{\mathcal{F}(Tx, s^{\star}) = accepted} \wedge \mathsf{\mathcal{F}(Tx, s) = accepted}
\end{numcases}.

\underline{$\expect{a1}$}: Naturally, the state $\mathsf{s}$ should be accepted, and $\mathsf{s^{\star}}$ will be abandoned within the time-bound $\delta$. If this event happens, $\mathcal{B}$ can break the promise of liveness. Thus, the probability of the event that fetches two different states from the blockchain is negligible. 

\underline{$\expect{a2}$}: Given a deterministic logic with the same initial state, the contract has accepted an invalid state. $\mathcal{B}$ has tampered the contract logic in some blockchain players, which contradicts the persistence requiring honest players to accept the contract logic in the deployment stage.

\underline{$\expect{a3}$}: A valid state $\mathsf{s}$ is not accepted, which contradicts \textit{liveness} since all valid states should be eventually confirmed within the time-bound $\delta$.

\underline{$\expect{a4}$}: The proof is similar to that in case (2). The contract returns an invalid state, indicating that the contract logic in some blockchain players has been altered, which contradicts the persistence assumption.

In a nutshell, the argument that $\mathcal{B}$ has fetched two different states contradicts our robust blockchain assumption. Thus, the adversary $\mathcal{A}$ can not win the game $\mathsf{G_{\adv,\mathsf{\Pi}}^{neqv}(\lambda)}$ with overwhelming probability. \qed
\end{prf}

\begin{thm}[Non-repudiation]
Assume that the blockchain is robust, our construction satisfies the property of non-repudiation.
\label{non-repudiation}
\end{thm}

\begin{prf}
The non-repudiation indicates that an adversary user can not deny having executed a certain function in a smart contract. If the transaction that invokes a contract is removed or changed, then we can build a new adversary $\mathcal{B}$ to break the assumption of the persistence property. Thus, the probability of $\mathcal{A}$ wining the game is negligible. We omit details since the proofs are the same as that in \underline{$\expect{a4}$} in the previous explanation.      \qed
\end{prf}

\begin{thm}[Non-frameability]
Assume that the blockchain is robust, and the signature scheme used in a transaction is secure against EUF-CMA, our construction satisfies the property of non-frameability.
\label{non-frameability}
\end{thm}

\begin{prf}
The non-frameability indicates that an adversary can not frame a honest user by producing the evidence of its ``misbehavior''. This evidence is usually represented as a transaction. Formally, we consider an adversary $\mathcal{A}$ and a challenger $\mathcal{C}$, and then define a sequence of games to finish this proof.

\textit{\textbf{Game $\Game_{0}$}}. This is an unmodified game. Trivially, the winning probability of this game equals the advantage of $\adv$ against non-frameability game, namely, $adv_{\mathcal{A}}^{\Game_{\textrm{nfrm}}}(\lambda)$.

\textit{\textbf{Game $\Game_{1}$}}. In this game, when the adversary $\mathcal{A}$ interacts with the challenger $\mathcal{C}$, $\mathcal{C}$ is not allowed to call the user oracle. Instead, $\mathcal{C}$ generates an invalid transaction $\mathsf{Tx}$ and returns it to $\mathcal{A}$.

\textit{\textbf{Game $\Game_{2}$}}. In this game, again, $\mathcal{A}$ is given an invalid transaction $\mathsf{Tx}$ to the contract. When $\mathcal{A}$ asks for executing a function in a contract, the smart contract stops execution and returns random state (e.g.,$\varnothing$) to $\mathcal{A}$.


Obviously, the winning probability of the game $\Game_{2}$, denoted as $\mathsf{adv_{\mathcal{A}}^{\Game_{\textrm{2}}}(\lambda)}$, is negligible, since the returned state is random selected. Next, to find out the differences between these games, we define some events. 

\underline{$\expect{a1}$: forging a transaction.} The event $\expect{a1}$ implies that the $\bdv_1$ forges a valid transaction $\mathsf{Tx}$ without interacting with the user oracle $\mathsf{O}^{\mathsf{user}}$ (denoted as $\neg\textbf{Sign}$), and $\mathsf{Tx}$ successfully triggers the executing the smart contract and the returned state successfully decrypts the ciphertext.


\begin{equation*}
\left.
\begin{array}{ll}
\begin{aligned}
       \mathsf{Tx} \gets \neg\mathsf{Sign(\sk_{tx},\mathsf{metadata}, aux)}.
       \\
       \mathsf{s} \xleftarrow{\mathbb{B}} \widehat{\mathcal{SC}}.\mathsf{Transfer} (\widehat{c}, \overline{\mathsf{s}},\mathsf{Tx}). \\
       \mathsf{ct} \xleftarrow{\;} \mathsf{Enc}(\mathsf{pk, s, m)} \\
       \mathsf{s^{\prime}} \xleftarrow{\mathbb{B}} \widehat{\mathcal{SC}}.\mathsf{Read}(\widehat{c}) \\
    \mathsf{m} \gets \mathsf{Dec}(\mathsf{sk, \mathsf{s^{\prime}} , ct}) \\
    \mathsf{true} \xleftarrow{\mathbb{B}} \mathsf{Inspect}(\mathsf{Tx})
\end{aligned}
\end{array}
\right]  \Rightarrow \expect{a2}.
\end{equation*}

\underline{$\expect{a2}$: adding an invalid transaction.} The event $\expect{a2}$ implies that $\bdv_2$ adds an invalid transaction to the blockchain, and the returned state successfully decrypts the ciphertext, denoted as $\neg\textbf{Transfer}$.

\begin{equation*}
\left.
\begin{array}{ll}
\begin{aligned}
       \mathsf{Tx} \gets \neg\mathsf{Sign(\sk_{tx},\mathsf{metadata}, aux)}.
       \\
       \mathsf{s} \xleftarrow{\mathbb{B}} \widehat{\mathcal{SC}}.\neg\mathsf{Transfer} (\widehat{c}, \overline{\mathsf{s}},\mathsf{Tx}). \\
       \mathsf{ct} \xleftarrow{\;} \mathsf{Enc}(\mathsf{pk, s, m)} \\
       \mathsf{s^{\prime}} \xleftarrow{\mathbb{B}} \widehat{\mathcal{SC}}.\mathsf{Read}(\widehat{c}) \\
    \mathsf{m} \gets \mathsf{Dec}(\mathsf{sk, \mathsf{s}^{\prime}, ct}) \\
     \mathsf{true} \xleftarrow{\mathbb{B}} \mathsf{Inspect}(\mathsf{Tx})
\end{aligned}
\end{array}
\right]  \Rightarrow \expect{a1}.
\end{equation*}

\textit{\textbf{Game $\Game_{0}$} $\approx$ \textbf{Game $\Game_{1}$}}. The winning condition for $\Game_{0}$ is equal to the condition for $\Game_{1}$ if and only if the event $ \expect{a1}$ does not happen. The probability of $\expect{a1}$ happening is equal to the advantage of breaking the promise of signature unforgeability (see EUF-CMA definition in Appendix A). Thus, we have

\begin{equation*}
     |  \prob{\Game_0} -
   \prob{\Game_1} | =
    \mathsf{\prob{\expect{a1}} = adv_{\bdv_1}^{\Game_{\textrm{EUF-CMA}}}}.
\end{equation*}

\textit{\textbf{Game $\Game_{1}$} $\approx$ \textbf{Game $\Game_{2}$}}.  In the deployment stage, the contract logic and contract initial state and the transaction that used to deploy such a contract are assumed to be confirmed, meaning that in  $\mathsf{\epsilon}$ blockchain players, the above data is stored more than $k$ blocks deep in their local ledgers. The \textbf{$\Game_{1}$} is the same with \textbf{ $\Game_{2}$}, since a contract should always reject an invalid transaction unless \textbf{$\Game_{1}$} accept an invalid transaction. Thus, the winning condition for $\Game_{1}$ is equal to the winning condition for $\Game_{2}$ if and only if the event $ \expect{a2}$ does not happen. The probability of $\expect{a2}$ happening is equal to the advantage of breaking the promise of persistence. Thus, we have

\begin{equation*}
     |  \prob{\Game_1} -
   \prob{\Game_2} | =
    \prob{\expect{a2}} = \mathsf{adv_{\mathcal{B}_2,(k,\delta)}^{persistence}}.
\end{equation*}

Putting everything together, we conclude that

\begin{equation*}
 \begin{aligned}
    adv_{\mathcal{A},{\mathsf{\Pi}}}^{\Game_{\textrm{nfrm}}}(\lambda) 
   \leq &
    \prob{\expect{a1}}
    +
    \prob{\expect{a2}}
    +
     \mathsf{adv_{\mathcal{B}}^{\Game_{2}}(\lambda)} \\
    \leq &
    \mathsf{adv_{\bdv_1}^{\Game_{\textrm{EUF-CMA}}}}
    +
    \mathsf{adv_{\mathcal{B}_2,(k,\delta)}^{persistence}}
    +
     \mathsf{adv_{\mathcal{A}}^{\Game_{2}}(\lambda)} 
    \leq 
     \mathsf{negl(\lambda)}.
 \end{aligned} 
\end{equation*}       \qed
\end{prf}

\section{Discussion}
\label{sec-discussion}
This section provides discussions on the benefits and potential challenges of smart contract-based hybrid (security) protocols. We find two widely-adopted usages of the smart contract applied in existing protocols: \textit{assisting TTP} and \textit{replacing TTP}, providing us with inspirations of application templates. Meanwhile, we explore its limitations, mainly, caused by underpinning blockchain systems: \textit{gas limit} and \textit{performance bottleneck}, warning us with negative instances. 

\subsection{Benefits} 
The hybrid protocols that combine the smart contract and security protocols bring benefits in two folds. On one side, a smart contract guarantees strong availability, the persistence of its state, and the correct execution of pre-defined protocols. On the other side, original security schemes provide secure, compact, and encryption/decryption functionalities. Our abstract properties of non-equivocation, non-repudiation and non-frameability have established the above claim. Then, we dive into a deep analysis of the improvements gain from additional (smart contract) benefits in each instance. 

In the first case, traditional CBE schemes rely on the certificate authority (CA). However, an illegal certificate revocation may disable the decryption capabilities of the corresponding certificate's owner. On the one side, CA may arbitrarily revoke a valid certificate, repudiate its actions, and indirectly make decryption fail. For example, Alice sends a ciphertext to Bob, but the evil CA has already revoked Bob's certificate without his permission. There is no way for Bob to obtain an up-to-date certificate, and, as a result, he cannot decrypt the ciphertext. On the other side, users cannot blame CA due to the absence of valid evidence on her malicious behaviour; worse still, the absence of incentive mechanisms for CA decreases their willingness to behave honestly. 

\begin{center}
\begin{tcolorbox}[colback=gray!10,
                  colframe=black,
                  width=12cm,
                  arc=1mm, auto outer arc,
                  boxrule=0.5pt,
                 ]
From this case (SC-CBE), \textit{we observe that smart contracts impact the security protocols by assisting traditional TTP. It supports the protocols by adding functions of validity checking, (honest) behaviours promoting, and data auditing.}
\end{tcolorbox}
\end{center}

Smart contract-based CBE protocols mitigate problems of malicious revocation, absent evidence and poor incentives. Firstly, the properties of non-equivocation and non-repudiation make the revocation conditions become publicly visible and checkable. Secondly, the smart contract receives revocation requests from users and then pushes valid ones to CA through transactions, where these transactions are used as evidence to detect illegal revocations, making CA and user's actions undeniable and accountable. Thirdly, our scheme automatically provides cryptocurrency-based rewards/punishments under predefined policies in the smart contract for CA's actions, which motivates CA to behave honestly.

In the second case, RBE \cite{RBE18,RBE19} has been proposed to solve the key escrow problem existing in the identity-based encryption~\cite{boneh2001identity} (IBE) scheme. It allows users to generate their own public and secret keys. However, RBE still places a significant amount of trust in its Key Curator (KC, a weak version of PKG), whose actions are not accountable. A dishonest KC may hide a trapdoor that enables to secretly create a key pair for yet unregistered identity or even register multiple keys for already registered users. The issues of strong centralization, untrusted KC, and identity authentication threaten the wide adoption. 

\begin{center}
\begin{tcolorbox}[colback=gray!10,
                  colframe=black,
                  width=12cm,
                  arc=1mm, auto outer arc,
                  boxrule=0.5pt,
                 ]
From this case (SC-RBE), \textit{we find that smart contracts can replace the traditional TTP by acting as an agent that takes over the tasks. It brings transparency and trust to centralized TTP.}
\end{tcolorbox}
\end{center}

In this case, we build a transparent RBE by coupling it with smart contracts. Smart contracts are used to automate the logic of KC transparently, enabling publicly upgradeable proofs on-chain that render KC's actions accountable. Equivalently, smart contacts take over the tasks of KC, replacing this centralized role with blockchain. The solution thereby transfers the right of key management from KC to users. A smart contract-based RBE delivers several advanced properties. Firstly, the scheme provides transparency to every participant, making KC's behaviours accountable. Users' registration and parameters can be publicly accessible. The only secret parameter is the user's private key generated by him/herself. This separation between key pair and public parameter effectively limits the Key Curator's ability to misbehave, and the permanent records on-chain make KC's actions traceable and accountable. Secondly, our scheme eliminates the reliance on a central authority and moves some related operations of KC on-chain, where the original KC is replaced by the smart contract with automatic state transitions. This replacement prevents external attacks, such as DDoS \cite{zargar2013survey} and MitM \cite{gelernter2017password}, through the robust blockchain system. Furthermore, the blockchain preserved by a group of players is rarely unavailable, guaranteeing that KC would always be online. 

\subsection{Challenges} 
Applying smart contracts to security protocols still confronts inevitable challenges. These issues mainly come from their inherent designs of supporting blockchain platforms.
To demonstrate such issues, we further implement a prototype\footnote{Proof of Concept implementation, source code avaliable at \url{https://github.com/typex2048/PKG-transparency/tree/master/service/target}} of the combined RBE construction to simulate the on-chain key accumulation process. The reason we implement RBE rather than CBE is due to the fact that RBE takes over the entire tasks of TTP, which is a perfect indicator to show the changes in terms of \textit{gas cost} and \textit{performance}.

\smallskip
\noindent\textbf{Gas consumption issue.}  Using a contract-based TTP for security protocols pays a high price on gas consumption. Under the initial design of Ethereum \cite{wood2014ethereum}, each contract activity related to calculation or storage, required a fee to be paid to the blockchain miners for providing computational resources. Furthermore, Ethereum sets a ceiling for the number of operations that can be included in one block. It is measured in the gas limits, whose upper bound reaches $12,134,453$ gas. In our implementation of the transparent RBE example, we mainly test \texttt{merge}, the most costly function in our protocol, which is used for key accumulation by leveraging the Merkel Tree. A single smart contract can support approximately $30$ times operations. Equivalently, it means the contract can stand up to around $14$ registered users ($28$ tree leaves,  $29$ times \textit{merge}). This is insufficient to support the wide adoption of decentralized applications. 

\smallskip
\noindent\textbf{Performance/scalability issue.} The security protocol that builds on the top of smart contracts also confronts performance and scalability bottlenecks. Smart contracts, running on distributed environments, have to communicate with peers for the validation of results and reach an agreement for the update of final states. This mechanism makes the combined system suffer from low performance and poor scalability issues. The combined system usually takes several seconds to thousands of seconds to complete a full cycle of TTP operations. This depends on the particular blockchain algorithm and network propagation. Worse still, neither increasing load of transactions nor increasing the number of nodes can improve the performance effectively.

\section{Conclusion}
\label{sec-conclu}
Smart contracts have been widely applied to security protocols. Few studies satisfactorily answer an intuitive question: \textit{how do smart contracts benefit these protocols}. In this paper, we try to give answers to such a question in depth. We firstly abstract a formal presentation of the smart contract. Then, we apply it into security protocols, presenting a generic framework with derived properties: \textit{non-equivocation}, \textit{non-repudiation} and \textit{non-frameability}. We further give two instances using our framework, separately  the CBE and RBE schemes, to highlight the roles (\textit{assist TTP} or \textit{replace TTP}). We provide strict security proofs based on formalized framework and proof-of-concept experiment results to demonstrate its feasibility and applicability to a variety of scenarios. To the best of our knowledge, this paper provides the first comprehensively theoretical discussions on applying smart contracts to security protocols. We believe that our formal treatment can provide an intuitive guide for proper combination between smart contracts and security protocols, as well as diversifying on-top DApps that rely on these protocols.

\normalem
\bibliographystyle{unsrt}
\bibliography{bib.bib} 
\section{Appendix A. Building Blocks}
\label{pri:se}
In this appendix, we provide the building blocks that are used for our concrete constructions of CBE and RBE schemes.

\smallskip
\noindent\textbf{Semantically Secure Encryption.}
A semantically secure encryption scheme $\mathsf{SE}$ consists of the following polynomial time
algorithms.

\begin{itemize}
\item[-] $ \mathsf{\mathsf{SE}.KeyGen}(1^\lambda)$ The probabilistic algorithm takes as input a security parameter $\lambda$ and outputs a key $\sk$.

\item[-] $ \mathsf{\mathsf{SE}.Enc(\sk,m)}$ The probabilistic algorithm takes as input $\sk$ and a message $\mathsf{m} \in \mathcal{M}$, outputs a ciphertext $\mathsf{ct}$.

\item[-] $ \mathsf{\mathsf{SE}.Dec(\sk,ct)}$ The deterministic algorithm takes as input $\sk$ and the ciphertext $\mathsf{ct}$, and outputs a message $\mathsf{m} \in \mathcal{M}$.

\end{itemize}

\smallskip
\noindent\textit{Correctness}. A semantically secure encryption scheme $\mathsf{SE}$ is correct if for all $\mathsf{m} \in \mathcal{M}$ and every key $\mathsf{\sk}$ outputted by $\mathsf{\mathsf{SE}.KeyGen(1^\lambda)}$, it holds that,
\begin{align*}
\mathsf{\mathsf{SE}.Dec(\sk,(\mathsf{SE}.Enc(\sk, m)))} = \mathsf{m}.
\end{align*}

A semantically secure encryption scheme provides data confidentiality. It should prevent an adversary from learning which message is encrypted in a ciphertext. Formally, the security of $\mathsf{SE}$ is defined as follows.

\begin{defi}[IND-CPA security of $\mathsf{SE}$]\label{secpa}
A semantically secure encryption scheme $\mathsf{SE}$ achieves Indistinguishability under Chosen-Plaintext Attack (IND-CPA) if
for all PPT adversaries, there exists a negligible function $\mathsf{negl(\lambda)}$ such that
\begin{align*}
\big| \Pr\big[ \mathsf{G_{\adv, \mathsf{SE}}^{IND-CPA}(\lambda)} = 1\big] - \frac{1}{2} \big| \leq \mathsf{negl(\lambda)},
\end{align*}
where $\mathsf{G_{\adv, \mathsf{SE}}^{IND-CPA}(\lambda)}$ is defined as follows. 

\begin{center}
 \procedure{$\mathsf{G_{\adv, \mathsf{SE}}^{IND-CPA}(\lambda)}$}{%
    \pcln \mathsf{\sk} \gets \mathsf{\mathsf{SE}.KGen}(1^\lambda) \text{//The challenger } \mathcal{C} \text{ runs this algorithm to obtain a key sk} \\
    \pcln \mathsf{b} \stackrel{\$}{\leftarrow} \{0,1\} \text{// } \mathcal{C}  \text{ chooses a random bit} \\
    \pcln \mathsf{(m_{0},m_{1})} \gets \mathcal{A}^{\mathsf{SE}(\cdot)} \text{//The adversary } \mathcal{A} \text{ provides a pair of messages } \mathsf{ (m_{0},m_{1})} \\
    \pcln \mathsf{ct^\star} \gets \mathsf{\mathsf{SE}.Enc(\sk,m_b}) \text{// } \mathcal{C}  \text{ replies
with } \mathsf{\mathsf{SE}.Enc(\sk,m_b}) \\
   \pcln \mathsf{b^{'}} \gets \adv^{\mathsf{SE}(\cdot)}\mathsf{(ct^\star)} \text{//} \mathcal{A}  \text{ finally outputs its guess }  \mathsf{b^{'}}\\
   \pcln \pcreturn  \mathsf{b = b^{'}}  } \\
 \end{center}

\end{defi}

\smallskip
\noindent \textbf{Signature Scheme.} 
A signature scheme $\mathsf{S}$ consists of the following algorithms.

\begin{itemize}
\item[-] $ \mathsf{\mathsf{S}.KeyGen}(1^\lambda)$ The algorithm takes as input a security parameter $\lambda$ and outputs a private signing key $\sk$ and a public verification key $\vk$.

\item[-] $  \mathsf{\mathsf{S}.Sign(\sk, m)}$ The algorithm takes as input $\sk$ and a message $\mathsf{m} \in \mathcal{M}$, and outputs a signature $\mathsf{\sigma}$.

\item[-] $ \mathsf{\mathsf{S}.Verify(\vk,\sigma,m)}$ The algorithm takes as input $\vk$, 
a signature $\mathsf{\sigma}$ and a message $\mathsf{m} \in \mathcal{M}$, and outputs $1$ or $0$.
\end{itemize}

\smallskip
\noindent\textit{Correctness}. A signature scheme $\mathsf{S}$ is correct if for all $\mathsf{m} \in \mathcal{M}$ and every key pairs $\mathsf{(\vk,\sk)}$ outputted from $\mathsf{\mathsf{S}.KeyGen(1^\lambda)}$, it holds that,

\begin{align*}
\mathsf{\mathsf{S}.Verify(\vk,(\mathsf{S}.Sign(\sk, m)),m)} = 1.
\end{align*}


A signature scheme provides authenticity. So an adversary without a signing key should not be able to generate a valid signature. The security of the signature scheme $\mathsf{S}$ is formally defined as follows. 


\begin{defi}[EUF-CMA security of $\mathsf{S}$]\label{secpa}
A signature scheme $\mathsf{S}$ is said to secure against Existentially Unforgeable under Chosen Message Attack (EUF-CMA) if for all PPT adversaries, there exists a negligible function $\mathsf{negl(\lambda)}$ such that 
\begin{align*}
 \Pr\big[ \mathsf{G_{\adv, \mathsf{S}}^{EUF-CMA}(\lambda)} = 1\big]  \leq \mathsf{negl(\lambda)},
\end{align*}
where $\mathsf{G_{\adv, \mathsf{S}}^{EUF-CMA}(\lambda)}$ is defined as follows.

\begin{pchstack}[center]
\begin{pcvstack}%
  \procedure{$\mathsf{G_{\adv, \mathsf{S}}^{EUF-CMA}(\lambda)}$}{%
    \pcln \mathsf{(\sk, pk)} \stackrel{}{\leftarrow} \mathsf{\mathsf{S}.KeyGen}(1^\lambda)  \\
    \pcln \mathcal{L} \gets \{\} \text{// an empty set}\\
    \pcln \mathsf{(m^{\star},\sigma^{\star})} \gets \mathcal{A}^{\mathcal{O}_{sign}( \cdot)} \mathsf{(m_0,m_1,\dots,m_n)}\\
   \pcln \pcreturn  \mathsf{ \mathsf{S}.Verify(vk,\sigma^{\star}, m^{\star})} = 1 \wedge \mathsf{\sigma^{\star} \notin \mathcal{L}}}
\end{pcvstack}
\pchspace
\procedure{$\mathcal{O}_{sign}( \cdot)$}{%
 \pcln \sigma \gets \mathsf{S.Sign(\sk, m)}  \\
 \pcln \mathcal{L} :=  \mathcal{L} \quad || \quad \sigma\\
 \pcln \pcreturn \sigma
 }
\end{pchstack}
\end{defi}

\smallskip
\noindent \textbf{Public Key Encryption.} 
A public key encryption scheme $\mathsf{PKE}$~\cite{cramer2007bounded} consists of the following algorithms.

\begin{itemize}
\item[-] $\mathsf{\mathsf{PKE}.KeyGen}(1^\lambda)$ The algorithm takes as input a security parameter $\lambda$ and generates a private key $\sk$ and a public key $\pk$.

\item[-] $ \mathsf{\mathsf{PKE}.Enc(\pk,m)}$ The algorithm takes as input a public key $\pk$ and a message $\mathsf{m} \in \mathcal{M}$, outputs a ciphertext $\mathsf{ct}$.

\item[-] $ \mathsf{\mathsf{PKE}.Dec(\sk,ct)}$ The algorithm takes as input a private key $\sk$ and a ciphertext $\mathsf{ct}$, and outputs a message $\mathsf{m} \in \mathcal{M}$.

\end{itemize}

\smallskip
\noindent\textit{Correctness}. A public key encryption scheme $\mathsf{PKE}$ is correct if for all $\mathsf{m} \in \mathcal{M}$ and all key pairs $\mathsf{(\sk,pk)}$ outputted from $\mathsf{ \mathsf{PKE}.KeyGen(1^\lambda)}$, it holds that, 

\begin{align*}
\mathsf{\mathsf{PKE}.Dec(\sk,(\mathsf{PKE}.Enc(pk,m))) = m},
\end{align*}

A public-key encryption scheme provides confidentiality. An adversary cannot learn which message is encrypted in a ciphertext, even if it equips with the decryption oracle before and after encryption. Formally, the security of $\mathsf{PKE}$ is defined as follows.

\begin{defi}[IND-CCA2 security of  $\mathsf{PKE}$]\label{ccapke}
A $\mathsf{PKE}$ scheme is said to secure against Indistinguishability Security Under Adaptively Chosen Ciphertext Attack (IND-CCA2) if for
all PPT adversaries, there exists a negligible function $\mathsf{negl(\lambda)}$ such that 
\begin{align*}
 \big| \Pr\big[ \mathsf{G_{\adv,\mathsf{PKE}}^{IND-CCA2}(\lambda)} = 1\big] - \frac{1}{2} \big|  \leq \mathsf{negl(\lambda)},
\end{align*}
where $\mathsf{G_{\adv,\mathsf{PKE}}^{IND-CCA2}(\lambda)}$ is defined as follows: 

\begin{center}
   \procedure{$\mathsf{G_{\adv, \mathsf{PKE}}^{IND-CCA2}(\lambda)}$}{%
    \pcln \mathsf{(\sk, pk)} \gets \mathsf{\mathsf{PKE}.KGen}(1^\lambda) \text{// } \mathcal{C} \text{ runs this algorithm to obtain } \mathsf{(\sk, pk)} \\
    \pcln \mathsf{(m_0,\cdots,m_n)} \gets \mathsf{\mathsf{PKE}.Dec}(\mathsf{ct_0,\cdots, ct_n}) \text{// } \mathcal{A} \text{ provides adaptively chosen } \mathsf{ct_{0,\cdots,n}} \\
    \pcln \mathsf{b} \stackrel{\$}{\leftarrow} \{0,1\}  \text{// } \mathcal{C} \text{ chooses a random bit}\\
    \pcln \mathsf{(m_{0},m_{1})} \gets \mathcal{A}^{\mathsf{PKE.Dec}(\sk,\cdot)} \text{// } \mathcal{A} \text{ provides } \mathsf{(m_{0},m_{1})} \text{ to } \mathcal{C}\\
    \pcln \mathsf{ct^\star} \gets \mathsf{\mathsf{PKE}.Enc(\sk,m_b}) \text{// } \mathcal{C} \text{ replies
with } \mathsf{\mathsf{PKE}.Enc(\sk,m_b}) \\
\pcln \text{// } \mathcal{A} \text{ continues to provide adaptively chosen } \mathsf{ct} \text{ with a restriction that } \mathsf{ct \neq ct^\star}\\
   \pcln \mathsf{b^{'}} \gets \adv^{\mathsf{\mathsf{PKE}.Dec}(\sk,\cdot)}\mathsf{(ct^\star)} \text{// } \mathcal{A} \text{ outputs its guess } \mathsf{b^{'}} \\
   \pcln \pcreturn  \mathsf{b = b^{'}}  } \\
\end{center}
\end{defi}

\smallskip
\noindent\textbf{Collision Resistant Hash Functions (CRHF)~\cite{bellare1997collision}} \label{pri:crhf} A family of functions $\mathsf{H} = \{\mathsf{h_0, h_1,\dots,h_k}\}$ is a collision resistant hash function family if each function $\mathsf{h_i}: \{0,1\}^{|m|} \to \{0,1\}^{|n|}, (i \leq k)$ satisfies the following properties:

\begin{itemize}
    \item[-] (length-compressing): $|m| > |n|$ 

    \item[-] (hard to find collisions): for any probabilistic polynomial algorithm $\mathsf{A}$, 
    \begin{align*} \Pr\big[\mathsf{ k \gets Gen(1^\lambda), (x1, x2) \gets A(k, 1^\lambda): x_1 \neq x_2 \wedge h_k(x_1) = h_k(x_2)}] \leq \mathsf{negl(\lambda)},
    \end{align*} 
\end{itemize}
where $ \mathsf{negl(\lambda)}$ is a \textit{negligible function}.

\smallskip
\noindent\textbf{Certificate-based encryption.} \label{pri:cbe} certificate-based encryption (CBE), firstly introduced by Gentry~\cite{gentry2003certificate}, has received considerable attention~\cite{galindo2008improved,liu2008efficient}. CBE is an intermediate paradigm that retains the desirable properties of public-key encryption and identity-based encryption~\cite{boneh2001identity}. In particular, it mitigates the certificate revocation problem by stopping the issuance of an implicit certificate for the revoked public key. A certificate-based encryption scheme $\mathsf{CBE}$~\cite{gentry2003certificate,galindo2008improved,liu2008efficient} consists of the following algorithms.

\begin{itemize}
\item[-] $\mathsf{\mathsf{CBE}.Gen(1^\lambda,n)}$ The algorithm takes as input a security parameter $\lambda$, the total number of time periods $\mathsf{n}$, and outputs the certifier’s master secret $\mathsf{msk}$ and public parameters $\mathsf{pms}$ that include the master public key $\mathsf{mpk}$.


\item[-]  $\mathsf{\mathsf{CBE}.Set(1^\lambda)}$ The algorithm takes as input $\lambda$, and outputs the user's key pair $\mathsf{(pk,sk)}$. The algorithm is run by users.

\item[-]  $\mathsf{\mathsf{CBE}.Cert(msk,i,user,pk)}$ At the start of each time period $\mathsf{i}$, CA takes as input $\mathsf{msk}$, user's information $\mathsf{user}$ and public key $\mathsf{pk}$, and outputs the certificate $\mathsf{Cert_{i}}$.

\item[-]  $\mathsf{\mathsf{CBE}.Enc(m,i,user,pk)}$ The algorithm takes as input $\mathsf{(m,user,pk)}$ at time period $\mathsf{i}$, and returns a ciphertext $\mathsf{ct}$ on message $\mathsf{m}$. 

\item[-]  $\mathsf{\mathsf{CBE}.Dec(Cert_{i},sk,ct)}$ The algorithm takes as input $\mathsf{(Cert_{i},sk,ct)}$ at time period $\mathsf{i}$, and then outputs a message $\mathsf{m}$ or a special symbol $\bot$ indicating a decryption failure. 
\end{itemize}

\smallskip
\noindent\textbf{Registration-based encryption.} \label{pri:rbe} registration-based encryption (RBE), proposed by Garg in 2018~\cite{RBE18,RBE19}, decouples the key generation process from the PKG altogether by replacing the PKG with a public key accumulator called Key Curator (KC). Every user in an RBE system generates its own public-secret key pair and sends the public key to the KC for registration. The KC is merely responsible for compressing all the registered user identity-key pairs into a short reference string. When a sender wants to send encrypted data to a receiver, he only requires the compressed public parameters along with the target identifier, whereas the receiver requires the public parameter along with the secret key associated with the registered public key. RBE consists of the following algorithms.

\begin{itemize}
\item[-] $ \mathsf{\mathsf{RBE}.Setup(1^{\lambda})}$ The algorithm takes as input a security parameter $\lambda$ and outputs a common random string $\mathsf{crs}$. 


\item[-] $ \mathsf{\mathsf{RBE}.KeyGen(1^{\lambda})}$ This algorithm takes as input a security parameter $\lambda$, and outputs a public key $\mathsf{pk}$ and a secret key $\mathsf{sk}$.

\item[-]  $\mathsf{\mathsf{RBE}.Reg^{[aux]}(\mathsf{crs,pp,id, pk})}$ The algorithm takes as input $\mathsf{crs}$, a public parameter $\mathsf{pp}$, a user identifier $\mathsf{id}$, a public key $\mathsf{pk}$ and outputs the updated public parameter $\mathsf{pp}'$.


\item[-] $\mathsf{\mathsf{RBE}.Enc(crs,pp, id, m)}$ The algorithm takes as input $\mathsf{crs}$, $\mathsf{pp}$, a recipient identity $\mathsf{id}$, a plaintext message $\mathsf{m}$, and outputs a ciphertext $\mathsf{ct}$.

\item[-] $\mathsf{\mathsf{RBE}.Upd^{[aux]}(pp, id)}$ The algorithm takes as input a user identifier $\mathsf{id}$, current public parameters $\mathsf{pp}$ and outputs newly updated public parameters $\mathsf{u}$ that can help $\mathsf{id}$ to decrypt its messages.

\item[-] $\mathsf{\mathsf{RBE}.Dec(sk, u, ct})$ The algorithm takes as input the secret key $\mathsf{sk}$, update information $\mathsf{u}$, a ciphertext $\mathsf{ct}$ and the outputs the message $m \in \{0, 1\}^*$ or in $\{\perp, \mathsf{GetUpd}\}$. The symbol $\perp$ indicates a syntax error, while $\mathsf{GetUpd}$ indicates that $\mathsf{u}$ needs to be updated.
\end{itemize}

\section*{Appendix B. Notations}
\label{appendix:B}

\begin{table*}[!hbt]
 \caption{Featured Notations}\label{tab-notation}
 \label{node}
  \centering
    \resizebox{\linewidth}{!}{  
    \begin{tabular}[t]{cll}
    \toprule
     \quad \textbf{Symbol} \quad  & \quad \textbf{Item}\quad   & \quad\quad\quad\quad\quad\quad\textbf{Functionalities} \\  \midrule
     $\mathsf{\Pi}$ & Hybrid protocols & smart contract-based security protocols  \\  \midrule 
     $\widehat{\mathcal{SC}}$ &  Smart contract & Implementing the transition of states on the top of blockchain platforms   \\  \midrule 
     $\widehat{c}$ & Contract & A piece of contract that contains the logic deployed from developers \\ 
     \midrule 
     $\mathcal{L}$  & Predefined logic & the contents containing the operation guidance written in the contract\\ \midrule
     $\Theta$  & Triggering condition & the conditions that triggers the execution of contracts\\  \midrule
     $f$  & Transition function & indicating the changes of states\\ \midrule
     $\mathcal{T}$  & Transaction \, & representing the set of transactions $\mathsf{Tx}$ in blockchain network \\  \midrule
     $\mathbb{B}$  &  Blockchain  &  providing distributed environments for all types of basic functions \\\midrule
     $\mathcal{S}$ & State & the current conditions/phase of the content of contracts, initiated as $\mathsf{s}$ \\ \midrule
     $\mathsf{player}$ & Player & representing blockchain maintainers who has real impact on consensus \\  \midrule
     $\overline{\mathsf{bytecode}} $ &  Byte code & a type of machine code as the intermediate state    \\  \midrule
     $\overline{\mathsf{opcode}} $  & Operation code &  specifying the operation set to be executed  \\ \midrule
     $\mathsf{metadata} $  & Meta-data &  the original data that has not been executed  \\ \midrule
     $\mathsf{pk/sk}$ & Key pair & the public key and private key to encryption/encryption \\ \midrule
     $\mathsf{msk}$ & Secret key &  the master secret key in CBE \\ \midrule
     $\mathsf{cert}$ & Certificate &  the certificates in CBE \\ \midrule
     $\mathsf{id}$ & Identities & the identification of users  \\ \midrule
     $\mathsf{ct}$ & Ciphertext & the encrypted messages under the private key  \\ \midrule
     $\mathsf{pp}$ & Parameters  &  public parameters in RBE \\ \midrule
     $\mathsf{Tree}$ & Trees  &  the trees that records identities and public keys of users in RBE \\ \midrule
     $\mathsf{crs}$ & Randomness & the common random string in RBE \\ \midrule
    $\mathcal{O}$  & Oracle  & an environment that can provide ideal functionalities \\ \midrule
    $\mathcal{A}$ & Adversary &  an adversary who has some ability to launch attacks \\ \midrule
    $\mathcal{C}$ & Challenger &  the role with abilities to response the requests from attacks \\ \midrule
    $\mathsf{negl(\lambda)}$ &  Negligible function & a function to show the negligible differences in security proofs \\ \midrule
    $\mathsf{\mathcal{F}}$  & Specified function & a type of function used for specific usages \\ \midrule
    $\mathsf{\lambda}$  & Security parameter & a type of parameter to adjust the security level of algorithms \\ \midrule
    $\Game$ & Game &  an experiment that show the game and operations in proofs \\ \midrule
    $\mathbb{E}$ & Event &  an event with predefined conditions used in proofs \\ 
    \bottomrule
    \end{tabular}
    }
\end{table*}

\end{document}